
%
%
\input harvmac

\def\Titleh#1#2{\nopagenumbers\abstractfont\hsize=\hstitle\rightline{#1}%
\vskip .5in\centerline{\titlefont #2}\abstractfont\vskip .5in\pageno=0}
%

\def\CTPa{\it Center for Theoretical Physics, Department of Physics,
      Texas A\&M University}
\def\CTPb{\it College Station, TX 77843-4242, USA}

\def\HARCa{\it Astroparticle Physics Group,
Houston Advanced Research Center (HARC)}
\def\HARCb{\it The Woodlands, TX 77381, USA}
\def\UAa{\it Department of Physics and Astronomy, The University of
Alabama}
\def\UAb{\it Box 870324, Tuscaloosa, AL 35487-0324, USA}

%
%
\def\ie{\hbox{\it i.e.}}     
\def\eg{\hbox{\it e.g.}}     

\def\h{{\textstyle{1\over2}}}
\def\q{{\textstyle{1\over4}}}
\def\tq{{\textstyle{3\over4}}}
\def\rt{{\textstyle{1\over\sqrt{2}}}}

\catcode`\@=11 

\def\lsim{\mathrel{\mathpalette\@versim<}}
\def\gsim{\mathrel{\mathpalette\@versim>}}
\def\@versim#1#2{\vcenter{\offinterlineskip
    \ialign{$\m@th#1\hfil##\hfil$\crcr#2\crcr\sim\crcr } }}
\def\boxit#1{\vbox{\hrule\hbox{\vrule\kern3pt
      \vbox{\kern3pt#1\kern3pt}\kern3pt\vrule}\hrule}}

\def\cl{\centerline}

\def\r#1{$\bf#1$}
\def\rb#1{$\bf\overline{#1}$}
\def\ty{{\widetilde Y}}
\def\t1{{\tilde 1}}
\def\ov{\overline}
\def\F{\widetilde F}
\def\Fb{\widetilde{\bar F}}

\def\JL{J.L. Lopez}
\def\DVN{D.V. Nanopoulos}

\def\GeV{\,{\rm GeV}}

\def\wt{\widetilde}

\def\NPB#1#2#3{Nucl. Phys. B {\bf#1} (19#2) #3}
\def\PLB#1#2#3{Phys. Lett. B {\bf#1} (19#2) #3}

\def\PRD#1#2#3{Phys. Rev. D {\bf#1} (19#2) #3}
\def\PRL#1#2#3{Phys. Rev. Lett. {\bf#1} (19#2) #3}
\def\PRT#1#2#3{Phys. Rep. {\bf#1} (19#2) #3}

\def\IJMP#1#2#3{Int. J. Mod. Phys. A {\bf#1} (19#2) #3}
\def\TAMU#1{Texas A \& M University preprint CTP-TAMU-#1}
%
%
\nref\books{See \eg, {\it String theory in four dimensions}, ed.
by M. Dine, (North-Holland, Amsterdam, 1988);
{\it Superstring construction}, ed. by A.N. Schellekens
(North-Holland, Amsterdam, 1989).}
\nref\ABK{I. Antoniadis, C. Bachas, and C. Kounnas, \NPB{289}{87}{87}.}
\nref\AB{I. Antoniadis and C. Bachas, \NPB{298}{88}{586}.}
\nref\KLT{H. Kawai, D.C. Lewellen, and S.-H. Tye, \PRL{57}{86}{1832};
\PRD{34}{86}{3794}; \NPB{288}{87}{1}.}
\nref\KLST{H. Kawai, D.C. Lewellen, J.A. Schwarz, and S.-H.H. Tye,
\NPB{299}{88}{431}.}
\nref\BDG{R. Bluhm, L. Dolan, and P. Goddard, \NPB{309}{88}{330}.}
\nref\Reiss{H. Dreiner, \JL, \DVN, and D. Reiss, \NPB{320}{89}{401}.}
\nref\KLN{S. Kalara, J. Lopez, and \DVN, \PLB{245}{90}{421},
\NPB{353}{91}{650}.}
\nref\GO{For a review see \eg, P. Goddard and D. Olive, \IJMP{1}{86}{303}.}
\nref\ELNa{J. Ellis, J. Lopez, and \DVN, \PLB{245}{90}{375}.}
\nref\FIQ{A. Font, L. Ib\'a\~nez, and F. Quevedo, \NPB{345}{90}{389}.}
\nref\Lewellen{D. Lewellen, \NPB{337}{90}{61}; J. A. Schwarz,
\PRD{42}{90}{1777}.}
\nref\Barr{S. Barr, \PLB{112}{82}{219}, \PRD{40}{89}{2457}; J. Derendinger,
J. Kim, and \DVN, \PLB{139}{84}{170}.}
\nref\AEHN{I. Antoniadis, J. Ellis, J. Hagelin, and \DVN, \PLB{194}{87}{231}.}
\nref\JHW{For a recent review see \eg, \JL\ and \DVN, \TAMU{76/91},
to appear in Proceedings of the 15th Johns Hopkins Workshop on Current Problems
in Particle Theory, August 1991.}
\nref\ALR{I. Antoniadis, G. Leontaris, and J. Rizos, \PLB{245}{90}{161}.}
\nref\SMCY{B. Greene, K.H. Kirklin, P.J. Miron, and G.G. Ross,
\PLB{180}{86}{69}; \NPB{278}{86}{667}; \NPB{292}{87}{606}; S. Kalara and
R.N. Mohapatra, \PRD{36}{87}{3474}; J. Ellis, K. Enqvist, \DVN, and K. Olive,
\NPB{297}{88}{103}; R. Arnowitt and P. Nath, \PRD{39}{89}{2006}.}
\nref\SMOrb{L. Ib\'a\~nez, H. Nilles, and F. Quevedo, \PLB{187}{87}{25};
L. Ib\'a\~nez, J. Mas, H. Nilles, and F. Quevedo, \NPB{301}{88}{157};
A. Font, L. Ib\'a\~nez, F. Quevedo, and A. Sierra, \NPB{331}{90}{421}.}
\nref\SMFFF{A. Faraggi, \DVN, and K. Yuan, \NPB{335}{90}{347};
A. Faraggi, \TAMU{49/91}; K. Yuan, PhD Thesis Texas A\&M University (1991).}
\nref\revamp{I. Antoniadis, J. Ellis, J. Hagelin, and \DVN,
\PLB{231}{89}{65}.}
\nref\AEHNa{I. Antoniadis, J. Ellis, J. Hagelin, and \DVN,
\PLB{205}{88}{459}.}
\nref\AEHNb{I. Antoniadis, J. Ellis, J. Hagelin, and \DVN,
\PLB{208}{88}{209}.}
\nref\PS{T.T. Burwick, R.K. Kaiser, and H.F. Muller, \NPB{362}{91}{232};
D. Bailin, E.K. Katechou, and A. Love, \IJMP{7}{92}{153}.}
\nref\decisive{J. L. Lopez and \DVN, \PLB{251}{90}{73} and \PLB{268}{91}{359}.}
\nref\cryptons{J. Ellis, J. Lopez, and \DVN, \PLB{247}{90}{257}.}
\nref\ELNpd{J. Ellis, \JL, and \DVN, \PLB{252}{90}{53};
G. Leontaris and K. Tamvakis, \PLB{260}{91}{333}.}
\nref\RT{J. Rizos and K. Tamvakis, \PLB{251}{90}{369};
I. Antoniadis, J. Rizos, and K. Tamvakis, Ecole Polytechnique preprint
CPTH-A092.1191 (1991).}
\nref\STAB{J. L. Lopez and \DVN, \PLB{256}{91}{150}; S. Kalara, \JL, and \DVN,
\PLB{275}{92}{304} and \TAMU{94/91}.}
\nref\neutrino{G. Leontaris, \PLB{207}{88}{447}; G. Leontaris and \DVN,
\PLB{212}{88}{327}; G. Leontaris and K. Tamvakis, \PLB{224}{89}{319};
S. Abel, \PLB{234}{90}{113}; I. Antoniadis, J. Rizos, and K. Tamvakis, Ecole
Polytechnique preprint CPTH-A140.0192 (1992).}
\nref\Lacaze{I. Antoniadis, J. Ellis, R. Lacaze, and \DVN, \PLB{268}{91}{188}.}
\nref\thresholds{S. Kalara, \JL, and \DVN, \PLB{269}{91}{84}.}
\nref\price{I. Antoniadis, J. Ellis, S. Kelley, and \DVN, \PLB{272}{91}{31}.}
\nref\EKNI{J. Ellis, S. Kelley and D. V.  Nanopoulos, \PLB{249}{90}{441}.}
\nref\ILR{L. Ib\'a\~nez, D. L\"ust, and G. Ross, \PLB{272}{91}{251};
L. Ib\'a\~nez and D. L\"ust, CERN preprint CERN-TH.6380 (1992).}
\nref\BLa{D. Bailin and A. Love, Sussex preprint SUSX-TH-91/16.}
\nref\BL{D. Bailin and A. Love, Sussex preprint SUSX-TH-91/17.}
\nref\SISM{S. Kelley, \JL, and \DVN, \TAMU{84/91} (to appear in Phys. Lett.
B).}
\nref\SS{N. Seiberg and E. Witten, \NPB{276}{86}{272};
L. Alvarez-Gaum\'e, G. Moore and C. Vafa, Comm. Math. Phys. {\bf 106}
(1986) 1.}
\nref\ABKW{I. Antoniadis, C. Bachas, C. Kounnas, and P. Windey,
\PLB{171}{86}{51}.}
\nref\GSO{F. Gliozzi, J. Scherk and D. Olive, \NPB{122}{77}{253}.}
\nref\SENE{D. Senechal, \PRD{39}{89}{3717}.}
\nref\DKV{L. Dixon, V. Kaplunovsky, and C. Vafa, \NPB{294}{87}{43};
T. Banks, L. Dixon, D. Friedan, and E. Martinec, \NPB{299}{88}{364}.}
\nref\LieI{See \eg, R. Gilmore, {\it Lie Groups, Lie Algebras, and
Some of Their Applications} (John Wiley \& Sons, New York, 1974).}
\nref\Hump{See \eg, J. Humphreys, {\it Introduction to Lie Algebras and
Representation Theory} (Springer Verlag, New York, 1980).}
\nref\LieII{R. Slansky, \PRT{79}{81}{1}.}
\nref\BG{J. Banks and H. Georgi, \PRD{14}{76}{1159}.}
\nref\Alon{A. Faraggi and \DVN, \TAMU{78/90} (unpublished).}
\nref\LN{\JL\ and \DVN, \NPB{338}{90}{73}.}
\nref\Kap{V. Kaplunovsky, \NPB{307}{88}{145}; J. Derendinger, S. Ferrara,
C. Kounnas, and F. Zwirner, CERN preprint CERN-TH.6004/91 (revised version).}
\nref\EVA{S. Kelley, \JL, and \DVN, \PLB{261}{91}{424}.}
\nref\flaton{J. Ellis, J. Hagelin, S. Kelley, \DVN, and K. Olive,
\PLB{209}{88}{283}.}
\nref\EKNIII{J. Ellis, S. Kelley, and \DVN, CERN preprint CERN-TH.6140/91.}
\nref\ENR{J. Ellis, \DVN, and D. Ross, \PLB{267}{91}{132}.}

\Titleh{\vbox{\baselineskip12pt\hbox{CTP--TAMU--11/92}\hbox{ACT--1/92}
\hbox{UAHEP922}}}
{\vbox{\cl{The Search for a Realistic Flipped SU(5) String Model}}}
\cl{JORGE L. LOPEZ$^{(a)(b)}$\footnote*{Supported
by an ICSC--World Laboratory Scholarship.},
D. V. NANOPOULOS$^{(a)(b)}$, and KAJIA YUAN$^{(c)}$}
\bigskip
\cl{$^{(a)}$\CTPa}
\cl{\CTPb}
\cl{$^{(b)}$\HARCa}
\cl{\HARCb}
\cl{$^{(c)}$\UAa}
\cl{\UAb}
\vskip .3in
\cl{ABSTRACT}
We present an extensive search for a general class of flipped $SU(5)$ models
built within the free fermionic formulation of the heterotic string. We
describe a set of algorithms which constitute the basis for a computer program
capable of generating systematically the massless spectrum and the
superpotential of all possible models within the class we consider. Our
search through the huge parameter space to be explored is simplified
considerably by the constraint of $N=1$ spacetime supersymmetry and the need
for extra $Q,\bar Q$ representations beyond the standard ones in order to
possibly achieve string gauge coupling unification at scales of
${\cal O}(10^{18}\GeV)$. Our results are remarkably simple and evidence the
large degree of redundancy in this kind of constructions. We find one model
with gauge group
$SU(5)\times U(1)_\ty\times SO(10)_h\times SU(4)_h\times U(1)^5$ and fairly
acceptable phenomenological properties. We study the $D$- and $F$-flatness
constraints and the symmetry breaking pattern in this model and conclude that
string gauge coupling unification is quite possible.
\bigskip
\Date{February, 1992}

\newsec{Introduction}
String model-building has provided a new and profuse source of ideas to
overcome many of the weaknesses of traditional unified models. Indeed, any
string-derived model possesses a definite gauge group whose gauge couplings
unify at a calculable high-energy scale, and its matter representations and
their gauge and Yukawa interactions are completely determined. In practice
this means that one faces a rather constrained problem with little or no room
for ways out of potentially phenomenologically disastrous situations.
Fortunately (or unfortunately) there is a very large number of models to choose
from and one then hopes that an educated path through the morass of
possibilities might lead to a model which describes the features of the
low-energy world and also predicts new observable phenomena. The purpose of
this paper is to pursue one such path through a (hopefully sizeable) portion
of this space.

There are several classes of constraints that one can apply to the universe
of possibilities to reduce the sample that needs to be considered. The most
basic ones are: $N=1$ spacetime supersymmetry and a gauge group below the
string scale that either
includes the standard model gauge group or embeds it but can be dynamically
broken down to it at a lower energy scale. There is also an array of
phenomenological constraints that need to be satisfied by the successful
model: the low-energy spectrum must contain the three generations of quarks
and leptons, and the very accurately measured values of $\sin^2\theta_w$,
$\alpha_e$, and $\alpha_3$ must be reproduced. The fulfillment of both these
constraints depends on the details of the path from the Planck scale down to
low energies, and are therefore harder to enforce given
our present lack of understanding of the mechanics of supersymmetry breaking.
At a deeper level of detail we must reproduce the observed values of the
quark and lepton masses. Any string-derived model which satisfies all these
constraints is a candidate for the correct theory of all particles and
interactions, and can be tested by \eg, their predictions for the top-quark,
neutrino, higgs, and sparticle masses.

{}From the string theory point of view there are some organizing principles.
It is widely believed that all string models can be thought of as points in
a parameter space called the moduli space. Each particular model having fixed
values of these parameters, although classes of models with continously
connected values of the parameters are very common. The hope is that eventually
a string theory principle will be found which will somehow select the
energetically most favorable point in this space. In practice, this space has
been parceled up into several different `formulations' of string theory \books\
which describe (not necessarily non-overlapping) subsets of models. For
model-building purposes we have chosen to explore (a portion of) this parameter
space with the visor of the so-called free fermionic formulation of the
heterotic string \refs{\ABK,\AB,\KLT,\KLST,\BDG,\Reiss}. This
formulation can be described as a set of model-building rules which is
amenable to systematic algebraic and symbolic manipulation and therefore can
be translated into a computer code. The description and implementation of such
code is one of the objectives of this paper. A further bonus of working
within this formulation is the ease with which the terms in the low-energy
effective action can be calculated \KLN, such as the cubic and higher-order
superpotential couplings.

String  theory also provides a correlation between the gauge group and its
allowed matter representations. First of all, these always come in anomaly-free
sets (except for a possible pseudo-anomalous $U_A(1)$ subgroup). The
four-dimensional gauge group is a reflection of the algebra of two-dimensional
world-sheet currents called the Kac-Moody algebra \GO. These algebras can be
realized at integer values of the `level' $k=1,2,\ldots$, all of which
represent the same gauge group. The allowed matter representations
depend on the chosen level which basically determines a cap on the
dimensionality of the allowed representations at that level \refs{\ELNa,\FIQ}.
By far the simplest and most common realizations
are level one. These allow a very limited set of matter representations which
do not include the adjoint representation. This result eliminates from further
consideration any of the traditional unified groups (\eg, $SU(5),SO(10),E_6$).
Higher-level realizations are possible (and they allow the adjoint
representation) although in practice these are harder to construct
\refs{\Lewellen,\FIQ}. Besides,
the unified groups that would become phenomenologically viable require high
levels of the Kac-Moody algebra to accommodate the large representations
which appear in traditional model-building with these gauge groups \ELNa.
Limiting ourselves to level-one Kac-Moody algebras we have two classes of
models to consider: the class of `flipped'-like $SU(n)$ models (of which
flipped $SU(5)$ \refs{\Barr,\AEHN,\JHW} is the simplest) and related models
\ALR, or models which contain explicitly the standard model gauge group
\refs{\SMCY,\SMOrb,\SMFFF} and therefore require no further symmetry
breaking. In this paper we explore the simplest unified models which can be
constructed using the simplest Kac-Moody algebras, \ie, the class of
flipped $SU(5)$ string models.

There already exists one string-derived realization of a flipped $SU(5)$ model
in this formulation,\foot{Flipped $SU(5)$ models have also been constructed
in other formulations \PS.} the so-called `revamped' flipped $SU(5)$ model
\refs{\revamp,\AEHNa,\AEHNb}. This model has been explored
in detail in the literature \refs{\decisive,\cryptons,\ELNpd,\RT,\STAB,
\neutrino} and several
of its interesting features have been highlighted, such as the natural
apperance of a hierarchical fermion mass spectrum \refs{\decisive,\RT},
adequate low-energy higgs sector \refs{\decisive,\RT}, acceptable proton decay
rate \ELNpd, the existence of bound states of fractionally charged hidden
sector particles called cryptons \cryptons, etc. Despite
all these attractive features, gauge coupling unification at a scale of
${\cal O}(10^{18}\GeV)$ as predicted in this model \refs{\Lacaze,\thresholds},
is {\it probably} not possible \price,
or conversely, the low-energy values of $\sin^2\theta_w$ and $\alpha_3$ are
{\it probably} not reproducible. In any event,
one must consider these statements from the proper perspective since it is
tacitly assumed that the matter content of the model does not differ from the
minimal one. This is not so in the true string model which possesses several
fields beyond the minimal ones at poorly determined intermediate mass scales.
Nevertheless, no well motivated choice for these mass scales has improved the
situation, and there even exists a no-go theorem to this effect (see below).

Two ways out this impasse have been proposed \EKNI: string threshold effects
on the gauge couplings could reduce the effective unification scale down to
${\cal O}(10^{16}\GeV)$ \refs{\ILR,\BLa}, or new matter fields could delay
unification until
${\cal O}(10^{18}\GeV)$ \refs{\price,\BL}. The first alternative is not viable
in models built in the free fermionic formulation since threshold corrections
always increase the effective unification scale
\refs{\thresholds,\ILR}.\foot{The threshold
corrections needed in models where this mechanism may work require rather
unnatural values of the moduli fields \ILR.} The minimal field-theoretical
extra matter content (beyond the fields in the supersymmetric standard
model) needed in the second alternative has been determined to
be an extra pair of $Q$ and $D^c$ vector-like representations \SISM, which fit
inside a \r{10} representation of $SU(5)$. However, in a more complicated
situation like for a string-derived flipped $SU(5)$ model, one can only assert
that one extra $Q,\bar Q$ pair is absolutely necessary.
The main purpose of this paper is to search for such models within
the chosen formulation.

The organization of this paper is as follows. In Sec. 2 we review the
fundamentals of the free fermionic formulation and rephrase its model-building
rules to suit our computational purposes. In Sec. 3 we describe our computer
algorithms. In Sec. 4 we explore the parameter space for the desired models,
and
in Sec. 5 we present the results of the extensive computer search. In Secs. 6
and 7 we describe in some detail the most promising model found, and in Sec. 8
we summarize our conclusions.
\newsec{The free fermionic formulation}
It has become well known that consistent classical string vacua
(\ie, ``string models") in low spacetime dimensions ($d<d_c=10$),
especially in four dimensions, can be constructed in a variety of
formulations \books. Different formulations correspond to different
choices for the two-dimensional world-sheet conformal field theory, so that
these extra two-dimensional degrees of freedom,
which may or may not admit a spacetime interpretation, together
with the two-dimensional conformal fields that describe the observed
four-dimensional spacetime, ensure the (super)conformal invariance
and modular invariance of the full string theory. In the free fermionic
formulation \refs{\ABK,\AB,\KLT,\KLST,\BDG,\Reiss}, all the internal
two-dimensional degrees of freedom are fermionized by utilizing only
free world-sheet fermionic fields. The stringy consistency
conditions are cast into a set of constraints on the spin-structures
\SS\ of all the world-sheet fermions (\ie, on the boundary conditions of these
fermions as they are parallell transported around noncontractible loops
on the world-sheet) and a set of constraints on the relative contributions of
different spin-structures to the string partition function.

The two-dimensional world-sheet fermions should provide the exact amount
of central charge for the Virasoro algebra in order to render the quantum
theory
free of conformal anomalies, both in the left- and right-moving sectors.
For the four-dimensional heterotic string\foot{We follow the convention
that left-movers are supersymmetric and right-movers are
non-supersymmetric.}, $c_L=9$ and $c_R=22$ are needed; these conditions fix
the total number of world-sheet fermions. In the light-cone gauge, in addition
to the two left-moving fermionic fields $\psi^\mu$ ($\mu=1,2$) which are the
superpartners of the two transverse string coordinates $X^\mu$, the internal
fermionic content consists of 18 left-moving and 44 right-moving real
fermions, respectively.
The world-sheet supersymmetry in the left-moving sector is nonlinearly
realized among the left-moving fermions which must transform as the adjoint
representation of a semi-simple Lie algebra of dimension $3(10-d)=18$ \ABKW.
To achieve $N=1$ spacetime supersymmetry, this algebra must be
$SU(2)^6$ \Reiss, and therefore we can group the left-moving fermions into
6 triplets ($\chi^\ell, y^\ell, \omega^\ell$) ($\ell=1,2,\ldots,6$),
each transforming as the adjoint representation of $SU(2)$,
with the following supercurrent
\eqn\Isc{T_F=\psi^\mu \partial_z X_\mu + \sum_{\ell=1}^6 \chi^\ell y^\ell
\omega^\ell,}
where $T_F$ must be periodic or antiperiodic on the world-sheet,
corresponding to spacetime fermions or spacetime bosons.

In addition to the requirement of (super)conformal invariance, the
most important constraints on the spin-structures of world-sheet
fermions are obtained by considering modular invariance. The basic
idea is to investigate the properties of the string partition and
correlation functions under modular transformations, and then search
for the solutions that accommodate modular invariance. In the free
fermionic formulation, besides the constraints that modular invariance
imposes on the one-loop string partition function, further constraints
can be obtained in two different approaches: (1) requiring modular
invariance and factorization of multiloop partition functions \refs{\ABK,\AB};
or (2) performing a ``physical sensible projection" from the space of
world-sheet states onto the subspace of physical states which ensures the
correct spin-statistics connection \KLT. These two approaches have been shown
to be entirely equivalent \refs{\KLST,\BDG}. It turns out that the results of
such analysis can be described as a set of model-building rules. We now
give a self-contained account of these rules.

The first set of rules restricts the spin-structure assignments of the
world-sheet fermions. Before stating these rules, it is helpful to
introduce the spin-structure vectors which specify unambiguously the
spin-structures of all world-sheet fermions. Consider a $g$-loop world-sheet
$\Sigma_g$; there are $2g$ noncontractible loops.  For each loop,
a spin-structure of a set of world-sheet fermions is given by
a orthogonal matrix representation of the first
homotopy group $\pi(\Sigma_g)$ which is non-abelian for $g\geq 2$.
In the approach of Ref. \KLT, since only the one-loop partition function is
relevant, one can just specify the boundary conditions of the
world-sheet fermions around the two noncontractible loops of the
torus, and they are mutually commuting. However, in the approach
of Refs.~\refs{\ABK,\AB}, on a two- or higher-loop world-sheet, the
analysis of multi-loop modular invariance and factorization
becomes a rather delicate affair, since non-commuting boundary
conditions are allowed \AB. Nevertheless, simplification can be
achieved by restricting oneself to the subset of mutually commuting
boundary conditions, which amounts to replacing $\pi(\Sigma_g)$ by its
abelianized version, \ie, the first homology group $H_1(\Sigma_g)$.
Under this restriction, even in the approach of Refs.~\refs{\ABK,\AB}
( which we follow closely in this paper) the spin-structure matrices
can be simultaneously diagonalized for all noncontractible loops.
As a result, for any noncontractible loop $v$, one can simply specify the
corresponding spin-structure by a spin-structure vector
\eqn\SPV{v=\bigl[v(f_1),\ldots,v(f_n)\bigr], \quad v(f) \in (-1,1],}
where $n=n_r+n_c$, $n_r$ and $n_c$ are the numbers of the real and
complex fermions. The component $v(f)$ represents the boundary
condition of fermion $f$ under parallel-transport around this loop
\eqn\DEV{f\rightarrow -e^{i\pi v(f)}f.}
It is convenient to introduce a sign for each spin-structure vector $v$ as
follows, $\delta_v=e^{i\pi v(\psi^\mu)}$. This sign distinguishes between
spacetime bosons ($\delta_v=+1$) and spacetime fermions ($\delta_v=-1$)
\refs{\ABK,\AB}. Clearly, one can choose to define the spin-structure vector
differently. In fact, the $W$-vector used in Ref.~\KLT\ and the $V$-vector used
in Ref.~\KLST\ are related to our $v$-vector by $v(f)=1-2W(f)=-2V(f)$.

It was shown in Ref.~\AB\ that if one only considers the
special spin-structure vectors whose components are rational numbers, then all
possible rational spin-structure vectors that give rise to consistent string
models must form a finite additive group $\Xi$, which can be generated by a
basis ${\cal B}=\{b_1, b_2, \ldots, b_p\}$ of a finite number of spin-structure
{\it basis} vectors. Each basis vector $b_i$ has order $N_{b_i}$, which is
defined as the smallest positive integer such that
\eqn\Nb{N_{b_i}b_i={\bf 0}\pmod{2}.}
Here ${\bf 0}$ is a special spin-structure vector which assigns
antiperiodic boundary conditions to all world-sheet fermions; we will refer to
it as the Neveu-Schwarz vector in the rest of this paper.
The basis ${\cal B}$ must obey the following constraints:

(A1) It can be chosen to be a canonical basis, such that
\eqn\AIi{\sum_{i=1}^p m_ib_i={\bf 0}\pmod{2}\Longleftrightarrow m_i
=0\pmod{N_{b_i}}\ \forall i.}
In terms of this canonical basis, an arbitrary spin-structure vector
$v\in \Xi$ is an integer linear combination of basis vectors reduced
in interval $(-1,1]$, which we express as
\eqn\AIii{v=\biggl[\sum_{i=1}^p v_ib_i\biggr],\quad v_i=0,1,2,\ldots,
         (N_{b_i}-1).}

(A2) One of the basis vectors, which we choose to be the first one
$b_1$, satisfies
\eqn\AII{{1\over 2}N_{b_1}b_1={\bf 1}\pmod{2}.}
Here ${\bf 1}$ is another special spin-structure vector which assigns
periodic boundary conditions to all world-sheet fermions; we will refer to
it as the Ramond vector for convenience. Although the general forms of such
$b_1$ vector have been tabulated \Reiss, we will choose $b_1={\bf 1}$ for
simplicity, hence $N_{\bf 1}=2$.

(A3) If $N_{b_ib_j}$ is the least common multiple of $N_{b_i}$
and $N_{b_j}$, then
\eqn\AIII{N_{b_ib_j}(b_i\cdot b_j)=0\pmod{4}.}

(A4) For any $b_i\in {\cal B}$,
\eqn\AIVi{N_{b_i}(b_i\cdot b_i)=\biggl\{\eqalign{&0\pmod{4}\quad
                               {\rm if}\ N_{b_i}\ {\rm is\ odd};\cr
          &0\pmod{8}\quad {\rm if}\ N_{b_i}\ {\rm is\ even}.\cr}}
However, if $N_{b_i}(b_i+{\bf 1})={\bf 0}$ (mod 4), then $b_i$ must
satisfy a stronger condition
\eqn\AIVii{N_{b_i}(b_i\cdot b_i)=N_{b_i}({\bf 1}\cdot{\bf 1})\pmod{16}.}

In \AIII--\AIVii, the dot product between two spin-structure vectors $u$
and $v$ is defined by\foot{This differs from the definition in
Refs.~\refs{\KLT,\KLST}
by a overall sign due to the different conventions used.}
\eqn\dot{u\cdot v=\biggl\{{1\over 2}\sum_{{\rm real}\atop {\rm left}}
                              +\sum_{{\rm complex}\atop {\rm left}}
                         -{1\over 2}\sum_{{\rm real}\atop {\rm right}}
             -\sum_{{\rm complex}\atop {\rm right}}\biggr\}u(f)v(f).}

(A5) The number of real fermions that are simultaneously periodic
in any given three basis vectors $b_i, b_j, b_k$ is even \KLST, \ie,
\eqn\AV{\sum_{\rm real} b_i(f)b_j(f)b_k(f)=0\pmod{2}\quad (1\leq
                i, j, k \leq p).}

(A6) The spin-structure corresponding to each $b_i$ should be an
automorphism of the Lie algebra defining the world-sheet supercurrent,
and all such automorphisms must commute with one another.
In the case of our interest, the world-sheet Lie algebra is $SU(2)^6$,
and the possible commuting sets of automorphisms have been tabulated in
Ref. \Reiss. This rule restricts in a rather intricate way the possible
sets of basis vectors which are allowed to coexist. We defer to the next
section a further discussion of this constraint.

Having reviewed the constraints of the spin-structure vectors, we now move on
to
the second set of rules which restricts the relative contributions of
different spin-structures to the one-loop string partition function. As
it was shown in Refs.~\refs{\AB,\ABK}, this set of rules can be given
in terms of some proper normalized coefficients which enter the
one-loop partition function and depend upon the spin-structures.
Since such one-loop spin-structure coefficients are pure phases \AB,
any direct computer implementation would involve complex numbers. In order to
avoid this unnecessary complication, we use a $k$-matrix notation similar to
that introduced in Refs.~\refs{\KLT,\KLST}. More precisely, for a given basis
${\cal B}=\{b_1, b_2, \ldots, b_p\}$, we define a $p\times p$ positive integer
matrix (which we call the $k$-matrix) in terms of the one-loop spin-structure
coefficients $C[{b_i\atop b_j}]$ associated with basis vectors
$b_i$ and $b_j$ (see Ref.~\AB), such that its element $k_{ij}$ is
given by\foot{Note: Despite its same name, our $k$-matrix definition differs
from that in Refs.~\refs{\KLT,\KLST}.}
\eqn\Dkm{C\biggl[{b_i\atop b_j}\biggr]=\delta_{b_i}e^{{2\pi i k_{ij}}
/N_{b_j}},\quad\ (1\leq k_{ij} \leq N_{b_j}).}

With the $k$-matrix, the second set of model-building rules can be
written as follows:

(B1) There is freedom in choosing $k_{11}$. In fact, we have
$1\leq k_{11} \leq N_{b_1}$ and
\eqn\kIIa{8k_{11}=N_{b_1}(b_1\cdot b_1)\pmod{4N_{b_1}},}
or
\eqn\kIIb{8k_{11}=N_{b_1}\bigl[{b_1\cdot b_1}-4b_1(\psi^\mu)\bigr]\pmod
{8N_{b_1}}.}
Since we take $b_1={\bf 1}$, Eq. \kIIa\ gives $k_{11}=1$ and Eq. \kIIb\ gives
$k_{11}=2$.

(B2) Besides $k_{11}$, all other diagonal elements $k_{ii}$ ($i\ge2$)
of the $k$-matrix are completely fixed by the corresponding $k_{i1}$,
\eqn\kki{8k_{ii}=N_{b_i}\bigl[4k_{i1}+{b_i\cdot b_i}+
{{\bf 1}\cdot {\bf 1}}-2(2-N_{b_1})b_i(\psi^\mu)\bigr]\pmod{8N_{b_i}}
\ (i\ge2).}
Note that the last term on the right side vanishes when $N_{b_1}=N_{\bf 1}=2$.

(B3) The elements below the diagonal $k_{ij}$ ($i>j$) and the corresponding
elements above the diagonal are related as follows (for $i\not=j$),
\eqn\kij{4\bigl(N_{b_i}k_{ij}+N_{b_j}k_{ji}\bigr)
+2N_{b_i}N_{b_j}\bigl[b_i(\psi^\mu)+b_j(\psi^\mu)\bigr]
-N_{b_i}N_{b_j}\bigl(b_i\cdot b_j\bigr)=0\pmod{4N_{b_i}N_{b_j}}.}

It is easy to see that a $k$-matrix is completely determined by
$k_{11}$ and the elements below the diagonal (or equivalently
the elements above the diagonal). Therefore, for a given basis
${\cal B}$, there are $2\prod_{i\not=j}g_{b_ib_j}$ distinct $k$-matrices
associated with it \AB, where $g_{b_ib_j}$ ($i\not=j$) is the greatest
common divisor of $N_{b_i}$ and $N_{b_j}$ (which gives the number of choices
for $k_{ij}$).

Given a pair $({\cal B}, k)$, subject to rules (A1)--(A6) and rules
(B1)--(B3) respectively, one can construct a consistent four-dimensional
heterotic string model. The total Hilbert space of the model is a direct
sum of the sub-Hilbert spaces of all the sectors in the model, each of
which corresponds to a spin-structure vector $v$ of form \AIii\ and is
refered to as the $v$-sector and sometimes denoted as ${\cal H}_v$.
Furthermore, the physical states in the
$v$-sector are determined by the generalized GSO-projections \GSO, which
take various forms \refs{\AB,\KLT}. Strictly speaking, in order to
perform the GSO-projections correctly, special care must be taken when
dealing with real fermions, because the zero modes of real fermions with
periodic boundary conditions contribute to the string partition function \KLST.
Of course, if all the left- and right-moving real world-sheet fermions in the
model can be separately complexified (\ie, described by left- and right-moving
complex fermions obtained by pairing up two real fermions that have the same
spin-structure) then the model is just a special case of a complex fermion
model, and there is no real subtlety involved.

In this paper, however, we would like to consider a more general class of
models, in which only $2n_L$ ($2n_L<20$) of the left- and $2n_R$ ($2n_R<44$) of
the right-moving real fermions are complexifiable in the usual sense. Clearly,
the two left-moving fermionic fields $\psi^\mu$ should always be treated as one
complex fermion. (Our choice of $S$-vector (described below)
allows complexification of the $\chi^\ell$ into three complex fermions \KLN.)
In addition, any remaining left- and right-moving real fermions appear in equal
numbers, and can be grouped in left-right pairs which share the same boundary
conditions (periodic or antiperiodic) for all basis vectors, effectively
representing a non-chiral conformal field theory, \eg, an Ising model \KLN.
This class of pseudo-complex fermion models is a subset of the general class of
real fermion models considered in Ref.~\KLST, and is precisely the class of
``real fermion models" discussed in Ref.~\KLN. Restricting ourselves to the
pseudo-complex fermion models, we can divide the world-sheet fermions
into three categories:
(1) $n_L$ left-moving complex fermions; (2) $n_R$ right-moving
complex fermions; (3) $n_I$ Ising models each formed from a ``residual"
left- and a ``residual" right-moving real fermion. The numbers of
these three types of world-sheet fermions are related by
\eqn\NMB{n_I=44-2n_R=20-2n_L.}
Accordingly, we can rearrange any spin-structure vector $v$ into three
parts: a $n_L$-dimensional vector $v^L$, a $n_R$-dimensional vector $v^R$ and a
$n_I$-dimensional vector $v^I$. For the particular class of basis vectors we
will consider in this paper, below we will see that rule (A6) implies
that the elements in $v^L$ and $v^I$ can only be either 1 or 0, whereas the
entries in $v^R$ can be rational numbers, which describe genuine right-moving
complex world-sheet fermions. For any complex world-sheet fermion $f$, one can
define the fermionic charge in the $v$-sector as \refs{\KLT,\AEHNa}
\eqn\Qch{Q_v(f)=F_v(f)+{1\over 2}v(f),}
where $F_v(f)$ is the fermion number and $v(f)$ the boundary condition.
Therefore, for left- and right-moving complex fermions, there are
the fermionic charge vectors $Q_v^L$ ($n_L$-dimensional) and
$Q_v^R$ ($n_R$-dimensional) respectively. For the real fermions
in left-right pairs, the concept of the fermionic charge is not
applicable. However, in the models under consideration, their
effect can be taken into account simply by introducing two operators
for each Ising model $f^I$: (1) a ``fermion number difference" operator
$\Delta F_v(f^I)$, which equals the fermion number of the left-moving
real fermion minus that of the corresponding right-moving fermion
when $v(f^I)=0$, but equals zero when $v(f^I)=1$; (2) a ``chirality"
operator $\Gamma_v(f^I)$ such that $\Gamma_v(f^I)=1$ if $v(f^I)=0$,
and $\Gamma_v(f^I)=\pm 1$ if $v(f^I)=1$. Collectively, we have two
well-defined $n_I$-dimensional vectors $\Delta F_v^I$ and
$\Gamma_v^I$.

In terms of $Q^L$, $Q^R$, $\Delta F^I$ and $\Gamma^I$, we can rewrite the
GSO-projections in a very convenient way. In fact, for all basis vectors
$b_i$ ($1\leq i\leq p$), the physical states in the $v$-sector must
satisfy
\eqn\GSoI{b_i^L\cdot Q_v^L-b_i^R\cdot Q_v^R+b_i^I\cdot {\Delta F_v^I}+
{1\over 2}b_i^I\cdot (\Gamma_v^I-{\bf 1}^I)
=v(\psi^\mu)+b_i(\psi^\mu)+2\sum_{j=1}^p
{k_{ij}v_j\over N_{b_j}}\pmod{2},}
where ${\bf 1}^I$ is the $n_I$-dimensional vector with all the entries
equal 1. When a particular string state of $v$-sector that satisfies
\GSoI\ is obtained, its quantum numbers (with which the corresponding
vertex operator can be constructed) are completely determined by the
three vectors $Q_v^L$, $Q_v^R$ and $\Gamma_v^I$ \KLN. More explicitly,
the first entry of $Q_v^L$ describes the spacetime spin of the state,
the next three entries give the charges under the three $U(1)$'s which
make up the conserved $U_J(1)$ current of the $N=2$ world-sheet
supersymmetry algebra, and the remaining entries ($n_L-4$) are the
charges under residual left-moving global $U(1)$ symmetries. The
right-moving fermion charge vector $Q_v^R$ encodes partially
\foot{It is quite often the case that a full matter representation
of the gauge group is formed by states that come from more than
one sector of the model.} the charges under the Kac-Moody currents and
therefore the gauge quantum numbers. In particular, the {\it rank} of
the total gauge group $G$ equals the dimension of $Q_v^R$, that is
\eqn\Rank{{\rm rank(G)}=n_R.}
Finally, the vector $\Gamma_v^I$ is related to the eigenvalues of the
order (disorder) operator for all the Ising models, which we denoted
collectively by a $n_I$-dimensional vector $\sigma_v^I$ with
\eqn\Sg{\sigma_v(f^I)=v(f^I)\Gamma_v(f^I).}

The mass of a string state in the $v$-sector is given by the following
formula \AB\
\eqnn\MfII
$$\eqalignno{M^2&=-{1\over 2}+{1\over 8}\biggl\{{1\over 2}\sum_{\rm
{real\atop left}}+\sum_{\rm {complex\atop left}}\biggr\}v(f)v(f)+
\sum_{\rm left}({\rm frequencies}),\cr
&=-1+{1\over 8}\biggl\{{1\over 2}\sum_{\rm
{real\atop right}}+\sum_{\rm {complex\atop right}}\biggr\}v(f)v(f)+
\sum_{\rm right}({\rm frequencies}).&\MfII\cr}$$
Where the oscillator frequencies for a fermion $f$ are
\eqn\RreqI{{1+v(f)\over 2}+{\rm integer},}
and if $f$ is complex, its complex conjugate $f^*$ has frequencies
\eqn\RreqII{{1-v(f)\over 2}+{\rm integer}.}
Also, the bosonic oscillators contribute integer frequencies.
\newsec{Description of the computer program}
One of the advantages of the free fermionic formulation is that
the rules for model-building are quite simple and can be computerized,
which makes a fairly systematical study of the string models built in
this formulation possible. We are aware of one such attempt in the
literature \SENE, which was based on the formulation of Ref.~\KLT.
However, in Ref.~\SENE, only complex fermion models were considered,
while the more interesting models with real fermions were not taken
into account. In order to investigate the pseudo-complex fermion models,
we have independently developed a computer program, based on the
formalism described in Sec. 2 and with an approach different from
that of Ref.~\SENE. Our computer software is written in FORTRAN.
The bulk of the program is devoted to the generation of the massless
spectrum of a pseudo-complex fermion model once the basis and the
consistent $k$-matrix of the model are given. In addition, our
program is also able to calculate the complete trilinear superpotential
and identify all the fermion Yukawa couplings of the model. In this section
we explain our method and discuss some algorithms involved.
\subsec{The basis and the $k$-matrix}
To obtain a basis ${\cal B}$ that obeys the rules (A1)--(A6),
we should first decide on the first basis vector $b_1$. An obvious choice that
we make throughout this paper is $b_1={\bf 1}$. As long as $b_1$ is fixed, rule
(A2) is not relevant, and the remaining task is to generate additional basis
vectors such that the other rules are satisfied. We choose the second basis
vector $b_2$ to be the ``supersymmetry generator" $S$, which allows the
existence of spacetime supersymmetry by giving rise to massless gravitinos.
Of the several possible forms of $S$ that are compatible with the choice
$b_1={\bf 1}$ \Reiss, we choose the simplest one, which is the only one
that has been used for model-building in the literature, namely
\eqn\SvI{S=(1\ 100\ 100\ 100\ 100\ 100\ 100\ :\ {\bf 0}_R).}
Here the first entry is reserved for the two transverse $\psi^\mu$
treated as a single complex fermion; the following 18 entries correspond
to the left-moving internal fermions treated here as 6 triplets
($\chi^\ell, y^\ell, \omega^\ell$); and the vector ${\bf 0}_R$ after
the colon, which separates the left-movers from the right-movers, simply
indicates that all right-moving fermions have antiperiodic boundary
conditions.

An immediate bonus of choice \SvI\ is that the possible forms for the
left-moving part of all other basis vectors can be rather easily
determined. In fact, if $S$ is given as in \SvI, then in order for
the basis vectors to satisfy rule (A6), the left-moving internal
fermions must have either periodic or antiperiodic boundary conditions \Reiss,
organized as 6 triplets ($\chi^\ell, y^\ell, \omega^\ell$) of real fermions.
Furthermore, in any basis vector $b_i$, the boundary conditions for these
fermions must satisfy \Reiss
\eqnn\AVI
$$\eqalignno{&b_i(\psi^\mu)=b_i(\chi^\ell)+b_i(y^\ell)+b_i(\omega^\ell)
\pmod{2},&\AVI\cr
&b_i(\psi^\mu), b_i(\chi^\ell), b_i(y^\ell), b_i(\omega^\ell)\ \in
\{0, 1\}\quad (1\leq \ell \leq 6).\cr}$$
Because of our choice of $S$, the otherwise rather intricate check of rule (A6)
can be done very easily in the class of models we consider.
In addition, the specific form of the vector $S$ also gives us some
useful information on the right-moving part of all other basis vectors.
{}From \SvI\ it is easy to see that only
$y^\ell$ and $\omega^\ell$ ($\ell=1,\ldots,6$) are available to serve
as the ``residual" left-moving real fermions (since the $\chi^\ell$ can always
be complexified). In order to obtain the pseudo-complex fermion models, we
need to allow equal numbers of left- and right-moving real fermions. Therefore,
we deduce that the first 12 entries of the right-moving part (corresponding to
12 real fermions denoted by $\bar y^\ell$ and $\bar\omega^\ell$) of all basis
vectors will be either 0 or 1. Furthermore, we can always treat the remaining
right-moving fermions as complex, \ie, the last 16 entries of each basis
vector can be rational numbers in the interval $(-1, 1]$.

Any test vector with the above general features has a chance of being
a viable basis vector. Once such a vector is generated (either by hand
or by computer) our program first checks whether or not it satisfies rules (A3)
and (A4), by calculating the dot product of this vector with all the other
basis vectors that have been previously entered into the basis. If the test
vector fails this step, the program discards it and repeats the same procedure
for another test vector, until a good candidate is found. At this point, the
program records the relevant information about this new basis vector, such as
its order and the associated dot products. This whole process can be repeated
as many times as one wishes and is controled by an interactive command. After
having obtained a set of several basis vectors, this set would constitute a
basis ${\cal B}$ if rule (A1) and (A5) are also satisfied. The algorithm for
checking rule (A1) is very simple. Our program first generates all
possible linear combinations of form \AIii\ with the set of basis vectors,
and then looks for the Neveu-Schwarz vector ${\bf 0}$, which should appear only
once, when all the ``coordinates" $v_i=0$, if the basis is canonical. As far
as rule (A5) is concerned, it can simply be checked using \AV. However, since
we are particularly interested in the pseudo-complex fermion models, we have
designated a special subprogram to check whether the model is complexifiable
as described in Sec. 2. Since all complexifiable models satisfy rule (A5),
once this is established we would have obtained an allowed basis ${\cal B}$
which would give rise to a pseudo-complex fermion model. It is worthwhile to
point out that in our approach we start with basis vectors of the general form
described above, and do not specify $n_L$, $n_R$ or $n_I$. These numbers are
determined by the program after the complexification procedure is successful.
The advantage of this approach is that our program can treat equally any
pseudo-complex fermion model with gauge group of rank between 16 and 22 (see
Eqs. \NMB\ and \Rank).

Having generated a basis ${\cal B}$, all consistent $k$-matrices can be
systematically obtained by using \Dkm--\kij. In practice, the number of
possibilities is prohibitively large and one has to resort to random
generation of these matrices (see Sec. 5). However, even in the random
approach there are still many distinct $k$-matrices, although they do not
necessarily lead to distinct or phenomenologically acceptable string models.
To avoid a large cosmological constant at the string scale, we are interested
only in models with spacetime supersymmetry. Furthemore, to obtain chiral
fermions this better be $N=1$ supersymmetry (as opposed to $N=2$ or 4 which are
also possible). Such models can be obtained if the basis ${\cal B}$
contains enough basis vectors to break the initial $N=4$ supersymmetry down
to $N=1$ \Reiss. However, since the GSO-projections relevant to
the massless gravitinos in the model (we need one and only one gravitino for
$N=1$) depend on the choice of $k$-matrix, only a subset of the $k$-matrices
will give $N=1$ supersymmetric models in the end. We have written a special
subprogram which calculates the massless gravitinos provided by the $S$-sector
immediately after a $k$-matrix is generated. In this way, the program acts on
only those selected $k$-matrices and discards all others. In Sec. 4.2 we will
show that given a specific basis, the constraints on the $k$-matrices imposed
by requiring $N=1$ supersymmetry can be worked out explicitly, in which case
we just use the explicit constraints to test for $N=1$ supersymmetry.
\subsec{The massless sectors}
To find the massless spectrum of a string model, we only need to
consider those sectors which could possibly contain massless states.
All we need are the basis ${\cal B}$ and the mass formula \MfII.
For an arbitrary $v$-sector, let us consider the following two quantities
\eqna\Mlr
$$\eqalignno{M_L&=\biggl\{{1\over 2}\sum_{\rm
{real\atop left}}+\sum_{\rm {complex\atop left}}\biggr\}v(f)v(f)
=v^L\cdot v^L+{1\over 2}v^I\cdot v^I,&\Mlr a\cr
M_R&=\biggl\{{1\over 2}\sum_{\rm
{real\atop right}}+\sum_{\rm {complex\atop right}}\biggr\}v(f)v(f)
=v^R\cdot v^R+{1\over 2}v^I\cdot v^I.&\Mlr b\cr}$$
In the models under consideration, since the left-moving part of
all basis vectors is made up of 0's and 1's, the left-moving part of
any spin-structure vector $v$ of form \AIii\ will only have 0 or 1
entries as well. Therefore, there are only two possible forms of massless
states in the left-moving part: the non-degenerate bosonic vacuum with a
single fermionic oscillator of frequency ${1\over 2}$ acting on it, and
the
degenerate fermionic vacuum with no oscillator at all. Accordingly, a
$v$-sector will provide massless states only if $M_L=0$ or $M_L=4$.

For the right-moving part, since the last 16 entries can be arbitrary rational
numbers in the interval $(-1, 1]$, a similar analysis is to certain extent
basis-dependent. Nevertheless, if we confine ourselves to models in which the
only rational numbers that appear in the basis vectors are $\pm{1\over 2}$,
then these would also be the only rational numbers in any vector $v$. As a
result, the possible right-moving fermionic frequencies are ${1\over 4}$,
${1\over 2}$ and ${3\over 4}$, and in order for a $v$-sector of such models to
provide massless states, it has to have $M_R=0,2,4,6,8$.
\subsec{The gauge group}
The massless vector states of a heterotic string model, which are created by
exciting the left-moving antiperiodic transverse fermions $\psi^\mu$
and the right-moving internal fermions, are the gauge bosons.
\foot{Gauge bosons which arise by exciting left-moving internal degrees of
freedom preclude the existence of chiral (massless) spacetime fermions if the
resulting gauge group is abelian (non-abelian) \DKV.} These come from massless
sectors with $M_L=0$ and transform as the adjoint representation of the gauge
group $G$ of the string model.
Since the right-moving complex fermions are the building blocks of the
two-dimensional Kac-Moody currents that underlie the four-dimensional
gauge symmetry, the gauge group $G$ of each pseudo-complex fermion model
can be systematically and unambiguously determined by studying the
fermionic charge vectors $Q_v^R$ for all the sectors with $M_L=0$.
First of all (as mentioned in Sec. 2), the number of right-moving
complex fermions $n_R$, which only depends on the basis ${\cal B}$,
gives the {\it rank} of $G$. Secondly but most importantly, the
$Q_v^R$ vectors which label the gauge bosons always constitute
the roots of the Lie algebra associated with the gauge group $G$ \KLT.

We start with the gauge bosons in the Neveu-Schwarz sector ({\bf 0}-sector)
which is always present in all models (see \AIii). Due to the existence of the
``residual" antiperiodic right-moving real fermions (not present in complex
fermion models), to get gauge bosons we can either excite two right-moving
complex fermions or excite one right-moving complex fermion and one such
``residual" right-moving real fermion (in both cases we have $M_R=0$, \ie,
`00' gauge bosons). In the first case we get right-moving fermionic charge
vectors of the form
\eqn\QoI{Q_{\bf 0}^R(f_i)=\pm\delta_{ij}\pm\delta_{ik}\quad (j\not=k),}
as well as some vanishing ones. In the second case, these vectors take the form
\eqn\QoII{Q_{\bf 0}^R(f_i)=\pm\delta_{ij}\quad.}
The Neveu-Schwarz sector is special in that the $Q_{\bf 0}^R$ vectors of the
gauge bosons in this sector themselves constitute the roots of a Lie algebra
associated with a subgroup of $G$ with of same rank, which we refer to as
the ``minimal" gauge group $G_m$ of the string model.
Clearly, for the simple models in which the
Neveu-Schwarz sector is the only sector giving gauge bosons, $G_m$ and
$G$ coincide. It is remarkable to see that, from \QoI\ and \QoII,
not only the $Q_{\bf 0}^R$ vectors correspond to the roots of $G_m$,
but their elements also specify precisely the coordinates of these
roots in the standard orthonormal basis \LieI. In addition, the forms of
\QoI\ and \QoII\ imply that the minimal gauge group $G_m$ in general
is a direct product of simple factors of $SO(2m)$, $SU(m+1)$ and
$SO(2m+1)$, as well as some $U(1)$ factors. Here, the non-simply-laced
groups $SO(2m+1)$ arise simply because the ``residual" right-moving
real fermions are allowed in pseudo-complex fermion models (see
\QoII). Since roots always appear in pairs of opposite sign, we only
need to consider the {\it positive roots}, \ie, the $Q_{\bf0}^R$
whose first nonvanishing entry is positive.
In the Neveu-Schwarz sector, our program first constructs the 462
positive roots of the Lie algebra $D_{22}$, which corresponds to the
primordial gauge symmetry $SO(44)$ of four-dimensional heterotic
string models built in the free fermionic formulation. The
positive roots of the minimal gauge group $G_m$ are then
obtained by performing the GSO-projections \GSoI\ which in this case reduce to
\eqn\GSog{-b_i^R\cdot Q_{\bf 0}^R+b_i^I\cdot {\Delta F_{\bf 0}^I}
=0\pmod{2}.}
Because the $Q_{\bf 0}^R$ vectors are $n_R$-dimensional, and in the
pseudo-complex fermion models we have $16\leq n_R\leq 22$,
only the first $n_R$ entries of the 462 positive roots of $D_{22}$
enter in \GSog. (When $n_R<22$, the first $n_R$ entries of some of these
positive roots are all zeros; such $n_R$-dimensional zero vectors are obviously
not positive, and our program discards them automatically.)
After the positive roots in the Neveu-Schwarz sector are found this
way, the program is instructed by an interactive command, to either
find the {\it simple roots} from these positive roots (according to the
algorithm which we will describe below) and therefore pin down the minimal
gauge group $G_m$ for some testing purposes, or to go on
searching for all the additional gauge bosons in other sectors and
finally find the total gauge group $G$ of the model.

Besides the Neveu-Schwarz sector, any $v$-sectors of the model with
$M_L=0$ could also provide gauge bosons. The most common gauge bosons of this
type have $M_R=8$ (\ie, `08' gauge bosons); in practice we have also found
models with `06' gauge bosons. The procedure to find
these additional gauge bosons is the same: we start from a set of
appropriate fermionic charge vectors $Q_v^R$ in each of such sectors,
keeping only those vectors whose first nonvanishing entry is positive, and
then find the subset which survives the GSO-projections.
In contrast with the Neveu-Schwarz sector, two differences are worth
mentioning. First, for the Neveu-Schwarz sector the GSO-projections
\GSog\ are independent of the $k$-matrix. So once a basis ${\cal B}$
is given, the minimal gauge group $G_m$ is completely determined
(prior to the generation of the $k$-matrices). For other sectors that
give gauge bosons, the corresponding GSO-projections become
\eqn\GSgg{-b_i^R\cdot Q_v^R+b_i^I\cdot {\Delta F_v^I}=2\sum_{j=1}^p
{k_{ij}v_j\over N_{b_j}}\pmod{2},}
which clearly depend on the $k$-matrix. As a result, the total gauge group
$G$ of the string model varies with different choices for the
$k$-matrix. Second, although the gauge bosons of the Neveu-Schwarz
sector give rise to the minimal gauge group $G_m$, the gauge bosons
from other sectors by themselves do not have this property.
Instead, the additional gauge bosons transform as certain
representations of the minimal group $G_m$, such that
all these representations of $G_m$ combine with the adjoint
representation coming from the Neveu-Schwarz sector to make up precisely
the adjoint representation of an enlarged group $G$.

Having found all the positive roots, to uniquely identify the gauge
group $G$, the next step is to find the simple roots and calculate the
Cartan matrix. The same idea was followed in Ref.~\SENE, where a
recursive procedure was used to find the simple roots according to
the property of their inner products with each other. Here we take
a simpler approach, which is based on the following definition of
the simple roots: {\it A simple root is a positive root that cannot
be written as the sum of two other positive roots} \LieI. Our
program first adds the positive roots together pairwise, then it
compares all the positive roots with these two-root pairs and finds
the simple roots. This way we end up with $n_{\rm srt}\leq n_R$
simple roots in some arbitrary order.

To determine explicitly the factors in the gauge group we proceed in two
steps. We first isolate a set of the simple roots which forms a subspace by
itself, \ie, that would give a block diagonal submatrix in the Cartan
matrix in a suitable basis. That is, $\alpha_i\cdot\alpha_j\not=0$ for
$i,j\in{\bf b}$, but $\alpha_i\cdot\alpha_k=0$ for $i\in{\bf b}$,
$k\not\in{\bf b}$. The number of roots in this block is the rank $r_i$ of the
sought after gauge group factor $G_i$. We then calculate the Cartan matrix
$C_{ij}$ of this block of simple roots. If $r_i=1$ and $C_{11}=0\,(2)$ then
$G_i=U(1)\,(SU(2))$. For $r_i>1$ the determination of $G_i$ can be made
unambigously from $r_i$ and ${\rm det}\,C$, as seen from Table I, and the
following additional considerations: for simplicity we identify $C_l$ with
$B_l$, although proper identification is possible;
we distinguish $B_7$ from $E_7$ by the fact that the Cartan matrix of
the former is not symmetric; and we take into account explictly the isomorphism
$A_3\approx D_3$. The above procedure is repeated until all simple roots are
exhausted and then a symbolic expression for the total gauge group is printed
out. In order to be able to identify the matter representations automatically,
it is necessary to reorder the simple roots in each block in a standard way
(such as the one given in Ref. \LieII). This is possible once the $G_i$ have
been identified. However, the algorithm is non-trivial and cumbersome and
will not be described here.

When $n_{\rm srt}<n_R$, the gauge group $G$ also contains
$(n_R-n_{\rm srt})$ $U(1)$ factors. In the models that we are
considering, the $U(1)$ gauge group can arise in two ways:
(1) If a single right-moving complex fermion has unique boundary
conditions so that no other right-moving complex fermions
have the same ones in all the basis vectors, then this single complex
fermion provides a $U(1)$ factor, and the corresponding $U(1)$
charge of a given representation is specified by the value of
its fermionic charge in that sector; (2) If a unitary group arises
through the embedding $SU(m)\times U(1)\subset SO(2m)$, then a $U(1)$
appears as a ``trace", and the $U(1)$ charge of a given
representation in this case is the sum of the fermionic charges over
all fermions in the trace. However, there are more complicated embeddings
for $U(1)$ factors, identification of which needs to be done on a non-automated
case-by-case basis.
\subsec{The matter states}
The massless matter states can come from massless sectors with $M_L=0$ or 4.
Similar to the case of massless gauge bosons, the gauge group properties of the
massless matter states are carried by their right-moving fermionic charge
vectors $Q_v^R$, which are the weights of the corresponding representations of
the Lie algebra associated with the gauge group $G$ \KLT. (As explained in
Sec. 2, the left-moving fermionic charge vectors $Q_v^L$ also contain useful
quantum numbers of the states \KLN.)

Since we only consider models with $N=1$ spacetime supersymmetry,
the matter states form $N=1$ supermultiplets. A full $N=1$
supermultiplet is provided by states from a $v$-sector and the corresponding
$(v+S)$-sector, so we always treat them together. The fermionic and scalar
components of an $N=1$ supermultiplet transform under the same representation
of the gauge group, but their left-moving quantum numbers are different. For
the fermionic components we choose to keep states with positive spacetime
helicity, \ie, $Q_v^L(\psi^\mu)={1\over 2}$. The correct left-moving fermionic
charges for the corresponding scalar components then satisfy\foot{This result
is a generalization of Eq. (37) in Ref. \KLN\ [with the following notational
relations: $Q_v^L(f_i)\leftrightarrow\alpha,\beta,\gamma$;
$Q_{grav}^L(f_i)\leftrightarrow\alpha_{12},\alpha_{34},\alpha_{56}$] for
the case of an arbitrary component of the gravitino vertex operator.}
\eqn\BFQ{\sum_{i=2,3,4}Q_v^L(f_i)Q_{grav}^L(f_i)={1\over 2},}
where $Q_{grav}^L$ is the fermionic charge vector of the (positive-helicity)
gravitino in the model, which our program has determined, and the sum runs over
the three complexified left-moving fermions constructed from the $\chi^\ell$.

The way in which we find the right-moving fermionic charge vectors
$Q_v^R$ for all massless matter states is essentially the same.
We first construct a set of appropriate fermionic charges vectors
$Q_v^R$ for each sector, and only keep those that survive the GSO-projections
\GSoI. Since the weights of Lie algebra representations do
not necessarily come in pairs, we have to consider both positive and negative
weights. To specify the irreducible
representations of the matter states, we only need to keep track of
the highest weights that appear in the end. Again, differently from
the approach of Ref.~\SENE, we find the highest weights by first converting all
weights into the corresponding Dynkin labels using the properly reordered
simple roots, and then keeping only those weights whose Dynkin labels are all
non-negative integers \LieII.

The identification of the transformation properties of the selected set of
weights under the various $G_i$ groups is done via a look-up table. That is,
the program knows the Dynkin labels of all allowed representations under all
gauge groups which can occur in a level-one Kac-Moody construction (a limited
set \ELNa). Looping over all $G_i$ we obtain a symbolic label for each state,
\eg, 10--1--1 for a \r{10} under $SU(5)$ and a singlet under other two $G_i$'s.
For later purposes, the multiplicity and conjugacy classes \refs{\LieII,\ELNa}
(see Sec. 3.5) of each representation are also calculated and saved. We also
determine the contribution to the anomaly of each $SU(N)$, $N\ge3$ subgroup
$G_i$ by
each of the states in the spectrum, as follows
\eqn\anoma{A_i=\sum_{R_j} a_i(R_j)\,{\rm dim}(R_j/G_i),}
where $a(R)=(N-3)!(N-2n)/(N-n-1)!(n-1)!$ for totally antisymmetric
representations (the only ones that can occur at level one \ELNa) of dimension
${\rm dim}(R)={{N\choose n}}$ \BG. In Eq. \anoma,
${\rm dim}(R_j/G_i)$ is the dimension of $R_j$ under the subgroups other than
$G_i$. Since all these anomalies must vanish, this constitutes a rather
valuable
check of the correctness of the program.
\subsec{The superpotential}
Once the full massless spectrum is available, the program computes the cubic
superpotential couplings using the techniques of Ref. \KLN. This is done by
forming all possible $\phi_1\phi_2\phi_3$ sets and determining whether such
coupling is invariant under the various two-dimensional world-sheet symmetries.
One of these is equivalent to four-dimensional gauge invariance. This is
checked
using the conjugacy classes determined earlier: under $G_i$ we must have
$c_i(\phi_1)+c_i(\phi_2)+c_i(\phi_3)=0\,{\rm mod}(d_i)$. For example, for
$SU(n)$, $c({\bf1})=0,c({\bf n})=1,c({\bf n(n-1)/2})=2$,
$c(\bar{\bf R})=-c({\bf R})$, and $d=n$; for
$SO(2n)$, $c({\bf1})=0,c({\bf n})=2,c({\bf2^{n-1}})=-1$, and $d=4$. We also
compute explicitly
the conformal field theory correlators for the various Ising models
(corresponding to each left-right real fermion pair). For cubic couplings these
can be tabulated explicitly using the following multiplication table
\eqn\mult{\bordermatrix{
&I&\sigma_+&\sigma_-&f\cr
I&I&\sigma_+&\sigma_-&f\cr
\sigma_+&\sigma_+&I&f&\sigma_-\cr
\sigma_-&\sigma_-&f&I&\sigma_+\cr
f&f&\sigma_-&\sigma_+&I\cr},}
where $\sigma_\pm$ are the eigenvalues of the chirality operator defined
in Eq. \Sg\ (also known as the order/disorder Ising model operators),
and $f$ represents the corresponding  left- or right-moving real fermion.
For example, the correlator $\vev{\sigma_+\sigma_-f}$ is nonvanishing because
$\sigma_+\sigma_-f\to ff\to I$, according to Eq. \mult.

For flipped $SU(5)$ models, among the various cubic couplings there are
some of the form: $10_a\cdot\bar 5\cdot\bar5$, $10_b\cdot 10_c\cdot5_d$,
$\bar5\cdot1\cdot5_e$. These are candidates for the up- and down-quark type
and lepton-type fermion Yukawa couplings. The code identifies these couplings
and determines whether it is possible to have non-vanishing top, bottom,
and tau Yukawa couplings at the cubic level, \ie, whether there are couplings
such that $10_a=10_b$ or $10_a=10_c$, and $5_d=5_e$.

A separate portion of the code determines the set of possibly non-vanishing
quartic and higher-order superpotential couplings according to the
(significantly more complicated) rules of Ref. \KLN. All we want to point out
here is that the multiplication table given above can be used to determine
whether a high-point Ising model correlator (corresponding to real
fermion degrees of freedom not involved in the picture changing operations)
vanishes. This test reduces considerably the number of terms that need to
scanned by hand. In fact, at the quartic level this program gives only
non-vanishing couplings.
\newsec{The space of flipped SU(5) models}
The computer program described in the previous section can be used to study
generic models built within the free fermionic formulation. As discussed in
the Introduction, we restrict ourselves here to models with the flipped $SU(5)$
group as the observable part of the total gauge group. This constraint is
rather weak once one realizes the size of the space to be explored. Instead of
doing an exhaustive search of all possible bases and $k$-matrices, we choose a
well motivated subset of the space of basis vectors (seven vectors) and
consider a general class of vectors `$\alpha$' which effect the breaking
$SO(10)\to SU(5)\times U(1)$, as discussed below. We also consider all the
possible $k$-matrices compatible with the bases under consideration.
\subsec{The choice of basis}
We study a basis of eight vectors
${\cal B}=\{{\bf1},S,b_1,b_2,b_3,b_4,b_5,\alpha\}$, the first seven of which
are given in Table II. The choice of $\bf1$ and $S$ was discussed in Secs. 2
and 3.1. The presence of vectors $b_1$ and $b_2$ is necessary
to reduce the spacetime supersymmetry from $N=4$ down to $N=1$. The precise
form of these vectors is basically fixed by self-consistency constraints
\refs{\Reiss,\Alon}. The presence of $b_3$ is necessary to enlarge the
massless spectrum and possibly accommodate three generations of chiral
fermions.
It has even been argued \Alon\ that the $b_1,b_2,b_3$ set is a unique one which
must be present in any three-generation model. Basis vectors $b_4$ and $b_5$
are needed to allow for symmetry-breaking Higgs representations in the massless
spectrum. Their precise forms have not been fully classified, although their
left-moving entries are constrained by self-consistency conditions (see Sec.
2).
Also, these vectors should not destroy or enhance the spacetime supersymmetry
of the model. In sum, given the form of $S$, the first five vectors in $\cal B$
are to a large extent ``written in stone",\foot{The other possible forms of
$S$ \Reiss\ have remained virtually unexplored in the literature.} whereas the
forms of $b_4$ and $b_5$ have been chosen as in the revamped model \revamp.

The vector $\alpha$ serves the purpose of breaking the resulting $SO(10)$ gauge
symmetry down to $SU(5)\times U(1)$, and must therefore contain rational
entries
which we take for simplicity to be $1\over2$. We study the following generic
form of the $\alpha$-vector
\eqn\alp{\alpha=(0\ 0y^1y^1\ \cdots\ 0y^6y^6\ :\
\bar y^1\bar y^2\bar y^3\bar y^4\bar y^5\bar y^6\
\bar w^1\bar w^2\bar w^3\bar w^4\bar w^5\bar w^6\
\h\h\h\h\h\ \h\h\h\ \h\h\h\h1100),}
where Eq. \AVI\ has been used to set $\alpha(w^\ell)=\alpha(y^\ell)$. The
choice
for the fixed entries was made to a large extent to follow the revamped
example.
With the choice of $b_{1,2,3,4,5}$, the complexification procedure indicates
that (depending on $\alpha$) the possible gauge groups will have rank 16--19.
The actual rank in excess of 16 depends on how many of the three $U(1)$
symmetries left unbroken by $b_{1,2,3,4,5}$ are also left unbroken by $\alpha$.
These symmetries are generated by the following complexified pairs of real
fermions, $U_a:\bar w^2\bar w^3$, $U_b:\bar y^1\bar w^6$, and
$U_c:\bar y^4\bar y^5$. It is convenient to divide up the possible forms of
$\alpha$ into eight classes, according to the number and type of $U_{a,b,c}$
symmetries that they break or do not break. This classification is given in
Table III. (The revamped model belongs to class 3a.)
\subsec{The constraints on the $k$-matrix}
Given the basis ${\cal B}$ with eight vectors, the corresponding $k$-matrix is
$8\times8$, but only the 28 lower-diagonal entries are independent (see Sec.
2).
As shown below, this leads to a very large number of possible $k$-matrices,
most of which may not necessarily
lead to phenomenologically interesting models. We know present three classes
of constraints on the $k$-matrix which guarantee $N=1$ spacetime supersymmetric
models, and allow the possibility of three generations of matter fields and
symmetry-breaking Higgs fields.

The GSO projections in Eqs. \GSoI\ can be simplified considerably for
potential gravitino states in the $S$-sector, as follows
\eqn\GSoII{b_i^L\cdot Q_S^L=1+b_i(\psi^\mu)+k_{i2}\pmod{2}.}
For $b_i=\alpha$ we have $\alpha^L\cdot Q_S^L=0$ (since $\alpha(\chi^\ell)=0$)
and $\alpha(\psi^\mu)=0$, therefore $k_{82}=1$. This constraint follows
solely from the assumed form of $\alpha$ in Eq. \alp. Analogously one can
show that $b_1^L\cdot Q_S^L=b_4^L\cdot Q_S^L$ and
$b_2^L\cdot Q_S^L=b_5^L\cdot Q_S^L$, and therefore $k_{62}=k_{32}$ and
$k_{72}=k_{42}$ if we want to preserve at least one gravitino. It can also be
shown that $k_{52}$ is determined as follows $k_{52}=1+k_{21}+k_{32}+k_{42}\
({\rm mod}\,2)$. If we pick the positive-helicity gravitino state, \ie,
$Q^L_S(\psi^\mu)=\h$, then the eight possibilities for $Q_S^L(\chi^{1,2},
\chi^{3,4},\chi^{5,6})$ follow uniquely from the values of $k_{32,42,52}$
as follows: $Q_S^L(\chi^{12,34,56})=-\h(-1)^{k_{32,42,52}}$.

We must also have representations left from the $b_{1,2,3}$-sectors. In this
case a special property of these vectors, namely $b_i\cdot b_j=b_i(\psi^\mu)
-\sum_{k=1}^5 b_i(\bar\psi^k)=-4$ (where $\bar\psi^\mu$ are the first five
right-moving complex fermions) allows one to deduce that
\eqn\bbb{b_2^L\cdot Q^L_{b_1}-b_2^R\cdot Q^R_{b_1}=
b_3^L\cdot Q^L_{b_1}-b_3^R\cdot Q^R_{b_1}
=Q_{b_1}^L(\psi^\mu)-\sum_{k=1}^5Q_{b_1}^R(\bar\psi^k).}
It can then be shown that we must have $k_{43}=k_{53}$ for the states in the
$b_1$-sector to remain. A similar exercise for the $b_{2,3}$-sectors yields
the constraint $k_{43}=k_{53}=k_{54}$. Further study of the effect of the
remaining GSO projections indicates that if $k_{43}=k_{53}=k_{54}=1\,(2)$,
then we will get \rb{16},\rb{16}' (\r{16},\r{16}') representations from each
of $b_{1,2,3}$.

An analogous argument applied to the $b_{4,5}$-sectors yields further
constraints on the $k$-matrix, as follows
\eqna\bff
$$\eqalignno{
&k_{61}=k_{63}\Rightarrow k_{64}=k_{65},\qquad
k_{61}\not=k_{63}\Rightarrow k_{64}\not=k_{65},&\bff a\cr
&k_{71}=k_{74}\Rightarrow k_{73}=k_{75},\qquad
k_{71}\not=k_{74}\Rightarrow k_{73}\not=k_{75}.&\bff b\cr}$$
\subsec{The possible models}
Given our choice of basis ${\cal B}=\{1,S,b_1,b_2,b_3,b_4,b_5,\alpha\}$,
it is convenient to determine the possible sets of $SU(5)$ representations
coming from the sectors ${\cal H}_{b_1,b_2,b_3,b_4,b_5}$ since these will
contribute the Ramond states of the observable spectrum. It is not hard to
show that the possible representations are as follows:
\eqna\IVi
$$\eqalignno{b_1,b_2,b_3&: \bar{\bf5}+{\bf10}\ {\rm or}\ 2\cdot{\bf10}\
{\rm or}\ 2\cdot\bar{\bf5},&\IVi a\cr
b_4,b_5&: {\bf5}+{\bf10}\ {\rm or}\ \bar{\bf5}+\ov{\bf10}\ {\rm or}\
{\bf10}+\ov{\bf10}\ {\rm or}\ {\bf5}+\bar{\bf5},&\IVi b\cr
b_4+b_5&: ({\bf5}+\bar{\bf5})_v\ {\rm or}\ 2\cdot{\bf5}\ {\rm or}\
2\cdot\bar{\bf5},&\IVi c\cr}$$
where the \r{5},\rb{5},\r{10},\rb{10} representations are pieces of
\r{16},\rb{16} representations of $SO(10)$, whereas
$({\bf5}+\bar{\bf5})_v={\bf10}$ of $SO(10)$. We can now form all possible
combinations of the above representations such that the total $SU(5)$
anomaly is zero,\foot{Otherwise a non-trivial mixture with the hidden sector
must occur. (States from the NS-sector contribute zero to the anomaly.)}
and $n_{10}\ge4$, $n_{\ov{10}}\ge1$. Five possibilities
arise: $n_{10}/n_{\ov{10}}:4/1,4/2,5/1,5/2,6/2$. (These go with
$n_5/n_{\bar 5}:2/5,2/4,1/5,1/4,0/4$.) Of these possibilities, only 4/1 and
5/2 models are likely to yield three light \r{10}'s (or equivalently
$n_g=n_{10}-n_{\ov{10}}=3$). The important point here is that the 4/1 and
5/2 models are obtained only when the $({\bf5}+\bar{\bf5})_v$ representations
are the ones contributed by ${\cal H}_{b4+b5}$. This requirement imposes
a constraint on the possible $\alpha$-vectors as follows.

Since the $SO(10)$ symmetry is not broken down to $SU(5)\times U(1)$ until
the $\alpha$-vector GSO projection is performed, whether we get
$({\bf5}+\bar{\bf5})_v$ or $2\cdot{\bf5}$ or $2\cdot\bar{\bf5}$ from the
$b_4+b_5$-sector depends only on this last GSO projection. Reverting to the
old notation for the GSO projection, the coefficient which matters is
$C[{b_4+b_5\atop \alpha}]=\pm1\,(\pm i)$ for
$(b_4+b_5)\cdot\alpha={\rm even}\,({\rm odd})$. By explicit calculation one
can show that if $C[{b_4+b_5\atop \alpha}]=\pm i$, then $({\bf5}+\bar{\bf5})_v$
will survive if $\alpha(y^2)=\alpha(y^3)$ ($y^2\bar y^2$ and $y^3\bar y^3$
form two Ising models), otherwise we get $2\cdot{\bf5}$ or $2\cdot\bar{\bf5}$.
Given the form of $\alpha$ in Eq. \alp, it follows that
$\alpha(w^2)=\alpha(w^3)$ and therefore $w^2w^3$ can be complexified and
the $U_a:\bar w^2\bar w^3$ symmetry remains unbroken. Analogously, one can
show that if $C[{b_4+b_5\atop \alpha}]=\pm1$, then $({\bf5}+\bar{\bf5})_v$
survives if $\alpha(y^2)\not=\alpha(y^3)$ and therefore
$\alpha(w^2)\not=\alpha(w^3)$ and $U_a$ is broken. In sum, if
$(b_4+b_5)\cdot\alpha={\rm even}\,({\rm odd})$ then $U_a$ will be broken
(unbroken). Now,
\eqn\balp{(b_4+b_5)\cdot\alpha=-1+\h\{
\alpha(y^2)-\alpha(\bar y^2)
+\alpha(y^3)-\alpha(\bar y^3)
+\alpha(w^1)-\alpha(\bar w^1)
+\alpha(w^4)-\alpha(\bar w^4)\}=-1,}
since $y^2\bar y^2,y^3\bar y^3,w^1\bar w^1,w^4\bar w^4$ always form Ising
models. Therefore $U_a$ must remain unbroken and from Table III we see that
only $\alpha$-vectors from classes 1,2a,2b, and 3a will respect this.
The universe of possibilities then consists of $\alpha$-vectors from these
classes and the set of distinct $k$-matrices compatible with each basis.
Furthermore, only 5/2 models will contain the extra pair of $Q,\bar Q$
representations required for gauge coupling unification.
\newsec{The computerized search}
Given the large number of possibilities, it is clear that one will benefit
from a statistical approach to model generation. However, even random
generation of a statistically significant number of $k$-matrices (for a fixed
$\alpha$) is prohibitive: a basis of $p$ vectors goes with an $p\times p$
$k$-matrix, whose $p(p-1)/2$ lower diagonal entries are independent. In our
case
$p=8$ and there are $2^{28}\approx268\,$ million possible $k$-matrices.
Fortunately the constraints on the $k$-matrix derived in Sec. 4.2 are very
powerful. Extensive random generation of $k$-matrices indicates that there are
further redundancy factors to be obtained, as follows:
\item{(i)}$N=1$ supersymmetry determines $k_{52,62,72,82}$ once $k_{21,32,42}$
are given. However, the eight choices for $k_{21,32,42}$ (which correspond
to the eight possible choices for the supersymmetry generator) give
equivalent models. We then get a $2^7$ reduction factor.
\item{(ii)} To obtain states from ${\cal H}_{b_1,b_2,b_3}$ we need to impose
$k_{43}=k_{53}=k_{54}=1,2$. If the model ends up being a 5/2 (or 2/5) model,
the specific choice just picks 5/2 over 2/5 (\ie, interchanges
${\bf10}\leftrightarrow\ov{\bf10}$). This gives a $2^3$ reduction factor.
\item{(iii)}There is no obvious analytic constraint on the entries
$k_{31,41,51}$, and heuristically we find that their values do
not affect the phenomenological properties of the models.\foot{They may even
give rise to equivalent models, as expected from the internal discrete
symmetries associated with the various Ising models.} We therefore fix them;
a $2^3$ reduction factor.
\item{(iv)}Finally, we have the analytical correlations among $k_{61,63,64,65}$
and $k_{71,73,74,75}$ to obtain states from ${\cal H}_{b4,b5}$. A reduction
factor of $2^2$.

The combined analytical and heuristical reduction factor is $2^{15}$ which
reduces the universe of possible $k$-matrices down to $2^{13}=8192$. This
number of $k$-matrices is in practice still quite large and makes random
generation of $k$-matrices belonging to this class practically necessary to
be able to explore several $\alpha$-vectors. {\it A word of caution}:
since the number of possibilities is finite, one must make sure that each
generated $k$-matrix is distinct from all previous ones.\foot{This is not
very important for small samples of $k$-matrices since these obey the
following probability distribution
$n_{distinct}=N(1-e^{-n_{trials}/N})\approx n_{trials}+{\cal
O}(n^2_{trials}/N)$
($N=8192$ in this case).} This is not hard to do; we simply identify each
$k$-matrix by the decimal equivalent of the 13-digit binary number formed by
its independent entries in a fixed order. In order to gain confidence on
statistical results (\eg, percentages of models with a given feature) obtained
by sampling $\lsim10\%$ of the possible models, we generated all 8192
$k$-matrices for one $\alpha$-vector and found the exact results to be in
complete agreement with those obtained statistically from smaller
samples.\foot{This run took $\approx40$ hours of cputime in a Silicon Graphics
340S computer. In comparison, the $2^{28}$ $k$-matrices would take $\approx153$
years of cputime!}

The $\alpha$-vectors which we consider introduce a large dimension in the
space of possible models. Each of these $\alpha$-vectors has 24 free real
fermion entries ($2^{24}$ possibilities), although these are highly
constrained.
Rule (A6) (see Eq. \AVI{}) requires $\alpha(y^\ell)=\alpha(w^\ell)$
for this type of $\alpha$-vectors (they all have $\alpha(\chi^\ell)=0$).
There are also six more
constraints on the 24 entries, one from rule (A4) and five from rule (A3)
(the constraint from $b_i={\bf1}$ follows from rule (A4), and that from $b_i=S$
is automatically satisfied for our choice of $\alpha$-vector).
Finally, the splitting of $\alpha$-vectors into classes (see Sec. 4.1)
introduces correlations among three pairs of entries. Therefore, there are
24--6--1--5--3=9 independent real entries in any given $\alpha$-vector, a total
of $2^9=512$ possibilities.

The program described in Sec. 3 is capable of producing a large amount of
output. It was therefore necessary to screen possible models, \ie, calculate
the whole spectrum and cubic Yukawa couplings, and keep only summary
information about each model (\eg, $k$-matrix number, gauge group,
$n_{10}/n_{\ov{10}}$, fermion Yukawa couplings, number and type of hidden
sector fields, etc.). With the $k$-matrix information it was then easy to
retrieve any particular model afterwards.

The results of our search with the general class of $\alpha$-vectors we
consider can be summarized as follows:
\item{(a)}{\it Class 3a}. There are two rank-17 gauge groups possible for any
$\alpha$-vector in this class (the $k$-matrix determines which one occurs):
$SU(5)\times SU(4)\times SU(2)^4\times U(1)^6$ and
$SU(5)\times SU(4)\times SO(10)\times U(1)^5$. We did not find
any 5/2 model with the $\alpha$-vectors explored and for all possible
choices of $k$-matrices. The revamped model (a 4/1 model) is a prototype of
models belonging to this class.
\item{(b)}{\it Classes 2a and 2b}. There are two rank-18 gauge groups possible
for each $\alpha$-vector: $SU(5)\times SU(4)\times SU(2)^4\times U(1)^7$ and
$SU(5)\times SU(4)\times SO(10)\times U(1)^6$. All 5/2 models found (these
occur 1/8 of the time; about 1000 per $\alpha$-vector choice) come with the
second gauge group. They also have the same number of massless fields (67),
and the same number and type of hidden matter fields (six \r{4},\rb{4} and
seven \r{6} of $SU(4)$, and three \r{10} of $SO(10)$). These models fall into
two subgroups: half of them have a 2/3/2 Yukawa set (2 up-quark--type,
3 down-quark--type, and 2 lepton--type cubic Yukawa couplings) and the other
half have a 1/3/2 Yukawa set.
\item{(c)}{\it Class 1}. There is a large number of rank-19 gauge groups
possible in this case, although they do not occur for all $\alpha$-vectors
(some are more `prolific' than others). The following is a list of the 13
gauge groups which we have been able to identify:
$$\eqalignno{
&SU(5)\times SU(4)\times
\cases{SU(2)^4\times U(1)^8\cr
        SU(3)^2\times SU(2)^2\times U(1)^6\cr
        SU(3)\times SU(2)^3\times U(1)^7\cr
        SO(10)\times U(1)^7\cr}\cr
&SU(5)\times SU(5)\times
\cases{SO(10)\times U(1)^6\cr
        E_6\times U(1)^5\cr}\cr
&SU(5)\times SU(6)\times
\cases{SU(2)^3\times U(1)^7\cr
        SU(3)^2\times SU(2)\times U(1)^5\cr}\cr
&SU(7)\times SU(4)\times
\cases{SU(2)^3\times U(1)^7\cr
        SU(3)\times SU(2)^2\times U(1)^6\cr}\cr
&SU(7)\times SU(6)\times
\cases{SU(2)^2\times U(1)^6\cr
        SU(3)\times SU(2)\times U(1)^5\cr}\cr
&SU(9)\times SO(10)\times U(1)^6\cr}$$
It is interesting to note that (contrary to naive expectations) the
addition of the $\alpha$-vector to effect the breaking down to flipped
$SU(5)$, can quite likely enlarge the final gauge group and leave no
such group after all. There are two kinds of $\alpha$-vectors in this class
which give 5/2 models:
\itemitem{(c1)} The `price'-like $\alpha$-vectors, one of which was used in
Ref. \price\ to construct the first 5/2 flipped $SU(5)$ model in the
literature.
In this case 5/2 models (which occur 1/4 of the time) always come with the
gauge group $SU(5)\times SU(5)\times SO(10)\times U(1)^6$ and {\it do not}
possess right-handed leptons in their massless spectra (they also have problems
with the anomalous $U_A(1)$ cancellation), and are therefore irrelevant for
phenomenological purposes.\foot{In Ref. \price\ the gauge group for the
model presented was incorrectly determined to be $SU(5)\times SU(4)\times
SO(10)\times U(1)^7$, and so were some of its massless representations. This
problem may have arisen because contributions of additional gauge bosons
belonging to new `08' sectors may have been overlooked.}
\itemitem{(c2)} The other type of $\alpha$-vectors give 5/2 models (1/8 of
the time) with gauge group $SU(5)\times SU(4)\times SO(10)\times U(1)^7$
and have no problems with right-handed leptons or cancellation of $U_A(1)$.
These models have the same hidden matter spectrum as those 5/2 models
belonging to classes 2a and 2b, and they also split into two subgroups:
half of them have 1/3/2 and the other half have 1/4/2 Yukawa sets.\foot{There
is a further subdivision inside each of these sets into two different kinds
of 1/3/2 and 1/4/2 Yukawa sets.}

\newsec{The phenomenology of the various models}
In the previous section we exhibited the list of possible 5/2 models found
under our assumptions about the basis vectors. Given the Yukawa couplings
of the various models one can easily convince oneself that only the models
that belong to classes 2a or 2b which have the 2/3/2 Yukawa set are of any
interest. Indeed, in models with only 1 up-quark--type Yukawa coupling
(\ie, 1/3/2 or 1/4/2) one must identify that coupling ($F_t\bar f\bar h$)
with the one for the top quark, since no other such coupling appears at
cubic, quartic, or quintic order in the superpotential. The problem is that
$F_t$ does not appear in terms of the form $F_t F h$ at cubic, quartic, or
quintic order, and therefore we do not expect a (sizeable) bottom Yukawa
coupling. We thus concentrate on the 2/3/2 models (in comparison, the
fairly successful revamped model is a 2/3/3 (although 4/1) model).
\subsec{The model}
The model we consider in some detail here belongs to class 2b (although all
2/3/2 models in classes 2a and 2b are equivalent to this one) and has
the following $\alpha$-vector
\eqn\VIi{\alpha=(00\ 000\ 000\ 000\ 000\ 000\ 011:\ 000001\ 011001\
\h\h\h\h\h\ \h\h\h\ \h\h\h\h1100),}
and $k$-matrix (this is the same one as the one for the revamped model,
although there are 512 such $k$-matrices that will give the same results)
\eqn\VIii{k=\pmatrix{
2 &1 &2 &2 &2 &2 &2 &1\cr
1 &1 &1 &1 &1 &1 &1 &4\cr
2 &2 &2 &2 &2 &2 &2 &1\cr
2 &2 &2 &2 &2 &2 &2 &1\cr
2 &2 &2 &2 &2 &2 &2 &4\cr
2 &2 &2 &2 &2 &2 &2 &1\cr
2 &2 &2 &2 &2 &2 &2 &3\cr
2 &1 &1 &1 &1 &1 &2 &3\cr}.}
We start by listing in Table IV the massless fields and their transformation
properties under the rank 18 gauge group $SU(5)\times U(1)_\ty\times SO(10)_h
\times SU(4)_h\times U(1)^5$. The cubic and quartic superpotential
terms are easily calculated \KLN. We obtain
\eqna\VIiii
$$\eqalignno{W_3=g\sqrt{2}\{
&F_0F_1h_1+F_2F_2h_2+F_4F_4h_1+F_4\bar f_5\bar h_{45}+F_3\bar f_3\bar h_3
                                +\bar f_2 l^c_2 h_2+\bar f_5 l^c_5 h_2\cr
&+\rt F_4\bar F_5\phi_3+\h F_4\bar F_4\Phi_0
                        +\bar F_4\bar F_4\bar h_1+\bar F_5\bar F_5\bar h_2\cr
&+(h_1\bar h_2\Phi_{12}+h_2\bar h_3\Phi_{23}+h_3\bar h_1\Phi_{31}
                                +h_3\bar h_{45}\bar\phi_{45}+{\rm h.c.})\cr
&+\h(\phi_{45}\bar\phi_{45}+\phi^+\bar\phi^++\phi^-\bar\phi^-
                        +\phi_i\bar\phi_i+h_{45}\bar h_{45})\Phi_3
                        +(\eta_1\bar\eta_2+\bar\eta_1\eta_2)\Phi_0\cr
&+(\phi_3\bar\phi_4+\bar\phi_3\phi_4)\Phi_5
                +(\Phi_{12}\Phi_{23}\Phi_{31}+\Phi_{12}\phi^+\phi^-
                +\Phi_{12}\phi_i\phi_i+{\rm h.c.})\cr
&+T_1T_1\Phi_{31}+T_3T_3\Phi_{31}\cr
&+D_6D_6\Phi_{23}+D_1D_2\bar\Phi_{23}+D_5D_5\bar\Phi_{23}+D_7D_7\bar\Phi_{31}
                                                        +D_3D_3\Phi_{31}\cr
&+\h D_5D_6\Phi_0+\rt D_5D_7\bar\phi_3\cr
&+\F_4\Fb_6\bar\Phi_{12}+\h F_3\Fb_4\Phi_0+\h F_2\Fb_5\Phi_3
                        +\F_6\Fb_4\phi^++\rt \F_5\Fb_4\phi_4\cr
&+\F_1\Fb_2D_5+\F_2\Fb_4l^c_2\},&\VIiii a\cr}$$
and
$$\eqalignno{W_4=
&F_2\bar f_2\bar h_{45}\bar\phi_4+F_3\bar F_4 D_4D_6+F_3\bar F_5 D_4 D_7\cr
&+l^c_3\Fb_3\Fb_6D_7+l^c_5\F_2\Fb_3\bar\phi_3
                        +\F_1\Fb_3(\phi^+\bar\phi_3+\bar\phi^-\phi_3)\cr
&+\Fb_3\Fb_5D_7\bar\phi^-+\F_2\F_5D_3\phi^-+\F_2\F_6D_3\phi_4+\F_5\Fb_1D_2D_7\cr
&+\F_5\Fb_2D_1D_7+\F_3\Fb_3D_3D_6+\F_4\Fb_3D_4D_7+\F_5\Fb_4D_5D_7.
                                                        &\VIiii b\cr}$$
Calculable coefficients \KLN\ have been omitted from $W_4$.
\subsec{F- and D-flatness}
To preserve unbroken supergravity at the string scale, the D-terms of all
$U(1)$ symmetries and all the F-terms must vanish, \ie,
\eqna\VIiv
$$\eqalignno{\vev{W}&=\vev{{\partial W\over\partial\phi_i}}=0,&\VIiv a\cr
        \vev{D_A}&=\sum_i q^i_A|\vev{\phi_i}|^2+\epsilon=0,&\VIiv b\cr
        \vev{D_a}&=\sum_i q^i_a|\vev{\phi_i}|^2=0,\quad a=1,2,3,4\,,&\VIiv c\cr
        \vev{D_\ty}&=\sum_i \ty_i|\vev{\phi_i}|^2=0,&\VIiv d\cr}$$
where $\epsilon=g^2{\rm Tr}\,U_A/192\pi^2$. In these formulas, $q^i_A$ is the
charge of the $i$-th field under the anomalous $U_A(1)$ \LN:
$U_A=(1/{\rm Tr}\,U_A)\sum_{i=1}^5[{\rm Tr}\,U_i]\,U_i$, $q^i_a$ are the
charges under the 5--1=4 orthogonal traceless linear combinations, and $\ty_i$
are the charges under (the anomaly-free) $U(1)_\ty$. We obtain
\eqn\VIv{U_A=(-36U_1-12U_2+24U_3-12U_5)/{\rm Tr}\,U_A,\qquad
{\rm Tr}\,U_A=46.4758,}
and therefore\foot{To recover the proper mass units we recall that we have set
$\kappa=\sqrt{8\pi}/M_{Pl}=1$ in Eq. \VIiv{b}.}
$\epsilon=(3.7\times g\times10^{17}\GeV)^2$. Note that ${\rm Tr}\,U_4=0$;
in fact, almost all singlets (except $\eta_{1,2},\bar\eta_{1,2}$) are neutral
under this $U(1)$ symmetry. The rotated
traceless $U(1)$'s (excluding $U_4$) can be written as follows (proper
normalization is understood, although not relevant in what follows)
\eqna\VIvi
$$\eqalignno{U_A&=-3U_1-U_2+2U_3-U_5,&\VIvi a\cr
        U_1'&=U_3+2U_5,&\VIvi b\cr
        U_2'&=U_1-3U_2,&\VIvi c\cr
        U_3'&=3U_1+U_2+4U_3-2U_5.&\VIvi d\cr}$$
The constraints on the singlet vevs obtained by solving Eqs. \VIiv{b,c} are
as follows
\eqna\VIvii
$$\eqalignno{&x_{45}=\epsilon/15-{1\over2}V^2_3,&\VIvii a\cr
        &x_+-x_-=\epsilon/15+{1\over2}V^2_3,&\VIvii b\cr
        &x_{31}-x_{23}=\epsilon/5,&\VIvii c\cr
        &\sum_{i=3,4}(x_i)+2(x_{23}+x_+-x_{12})=\ov V^2_5-V^2_2,&\VIvii d\cr}$$
where $x_p\equiv|\vev{\phi_p}|^2-|\vev{\bar\phi_p}|^2$, and
$V_i\equiv\vev{\nu^c_i}$, $\ov V_i\equiv\vev{\bar\nu^c_i}$. Similarly,
$D$-flatness of the anomaly-free $U_4(1)$ and $U(1)_\ty$ requires
\eqna\VIviia
$$\eqalignno{|V_0|^2&=|V_1|^2+2(|\eta_1|^2-|\bar\eta_1|^2
                                +|\eta_2|^2-|\bar\eta_2|^2),&\VIviia a\cr
V^2&=\sum_{i=0}^4|V_i|^2=|\ov V_4|^2+|\ov V_5|^2=\ov V^2.&\VIvii b\cr}$$

The F-flatness constraints (Eqs. \VIiv{a}) can also be worked out for the cubic
superpotential and the resulting 24 equations (one for each singlet field) can
be summarized as follows
\eqna\VIviii
$$\eqalignno{&\Phi_{12}\Phi_{31}=\Phi_{12}\Phi_{23}=\phi_{45}\Phi_3=0\qquad
                                        {\rm and\ h.c.},&\VIviii a\cr
&\Phi_{23}\Phi_{31}+\phi^+\phi^-+\phi_i\phi_i=0\qquad {\rm and\ h.c.},
                                                        &\VIviii b\cr
&\phi_{45}\bar\phi_{45}+\phi_i\bar\phi_i+\phi^+\bar\phi^++\phi^-\bar\phi^-=0,
                                                        &\VIviii c\cr
&\phi_3\bar\phi_4+\bar\phi_3\phi_4=0,&\VIviii d\cr
&\phi^+\Phi_3+\bar\phi^-\bar\Phi_{12}=\phi^-\Phi_3+\bar\phi^+\bar\Phi_{12}=0
                                        \qquad{\rm and\ h.c.},&\VIviii e\cr
&\h V_4\ov V_4+\eta_1\bar\eta_2+\bar\eta_1\bar\eta_2=0,&\VIviii f\cr
&\rt V_4\ov V_5+\bar\phi_3\Phi_3+\bar\phi_4\Phi_5+2\phi_3\Phi_{12}=
        \phi_3\Phi_3+\phi_4\Phi_5+2\bar\phi_3\bar\Phi_{12}=0,&\VIviii g\cr
&\bar\phi_4\Phi_3+\bar\phi_3\Phi_5+2\bar\phi_4\bar\Phi_{12}=0
                                        \qquad{\rm and\ h.c.},&\VIviii h\cr
&\{\eta_1,\bar\eta_1,\eta_2,\bar\eta_2\}\Phi_0=0.&\VIviii i\cr}$$
These equations have five classes of solutions when solved in conjunction
with the D-flatness constraints in Eqs. \VIvii{}. There are eight possible
solutions to Eqs. \VIviii{a}, \ie, $\Phi_{12}=\bar\Phi_{12}=\Phi_3=0$,
$\Phi_{12}=\bar\Phi_{31}=\bar\Phi_{23}=\Phi_3=0$, $\Phi_{12}=\bar\Phi_{12}=
\phi_{45}=\bar\phi_{45}=0,\cdots$. Four of these solutions violate either of
the D-flatness conditions in Eqs. \VIvii{b,c}. The remaining five solutions
can be worked out in detail and are as follows:
\item{1.} $\Phi_{12}=\bar\Phi_{12}=\Phi_3=0$,
\itemitem{a.} $\Phi_5=0$, $V_4=0$, \VIvii{}, \VIviii{b,c,d,f};
\itemitem{b.} $\Phi_5=0$, $V_4\not=0,\ov V_5=0$, \VIvii{}, \VIviii{b,c,d,f};
\itemitem{c.} $\Phi_5\not=0$, $V_4=0$, $\phi_{3,4}=\bar\phi_{3,4}=0$, \VIvii{},
\VIviii{b,c,f};
\itemitem{d.} $\Phi_5\not=0$, $V_4\not=0$,
$\phi_{3,4}=\bar\phi_3=0$, \VIvii{}, \VIviii{b,c,f,g}.
\item{2.}$\Phi_{12}=\bar\Phi_{31}=\bar\Phi_{23}=\Phi_3=0$,
$\bar\phi^+=\bar\phi^-=0$, $\bar\phi_{3,4}=0$, $\phi_{45}\bar\phi_{45}=0$,
$V_4=0$, $\ov V^2_5>V^2_2$,
\itemitem{a.}$\Phi_5=0$, \VIvii{}, \VIviii{b,f};
\itemitem{b.}$\Phi_5\not=0$, $\phi_{3,4}=0$, \VIvii{}, \VIviii{b,f}.
\item{3.}$\bar\Phi_{12}=\Phi_{31}=\Phi_{23}=\Phi_3=0$, $\phi^+=\phi^-=0$,
$\phi_{3,4}=0$, $\phi_{45}\bar\phi_{45}=0$, $\ov V^2_5<V^2_2$,
\itemitem{a.}$\Phi_5=0$, \VIvii{}, \VIviii{b,f,g};
\itemitem{b.}$\Phi_5\not=0$, $\bar\phi_{3}=0$, \VIvii{}, \VIviii{b,f,g}.
\item{4.}$\Phi_{12}=\bar\Phi_{31}=\bar\Phi_{23}=0$,
$\phi_{45}=\bar\phi_{45}=0$,
$\bar\phi^+=\bar\phi^-=0$, $V^2_3=2\epsilon/15$,
\itemitem{a.}$\Phi_5=0$, $\phi_{3,4}=\bar\phi_{3,4}=0$,
$\ov V^2_5>V^2_2$, \VIvii{}, \VIviii{b,f,g};
\itemitem{b.}$\Phi_5\not=0$, \VIvii{}, \VIviii{b,c,d,f,g,h}.
\item{5.}$\bar\Phi_{12}=\Phi_{31}=\Phi_{23}=0$, $\phi_{45}=\bar\phi_{45}=0$,
$\phi^+=\phi^-=0$, $V^2_3=2\epsilon/15$,
\itemitem{a.}$\Phi_5=0$, $\phi_{3,4}=\bar\phi_{4}=0$,
$\ov V^2_5<V^2_2$, \VIvii{}, \VIviii{b,f,g};
\itemitem{b.}$\Phi_5\not=0$, \VIvii{}, \VIviii{b,c,d,f,g,h}.

\noindent In cases 2 and 3, the solution to $\phi_{45}\bar\phi_{45}=0$ depends
on the sign of $\xi=\epsilon/15-{1\over2}V^2_3$. For $\xi>0\,(<0)$,
$\bar\phi_{45}=0\,(\phi_{45}=0)$, whereas for $\xi=0$,
$\phi_{45}=\bar\phi_{45}=0$. Note also that cases 2 and 3 are mutually
exclusive. In solving Eqs. \VIviii{i}, we have assumed that
$\eta_{1,2},\bar\eta_{1,2}$ have generically nonvanishing vevs and therefore
$\Phi_0=0$. If $\eta_{1,2}=\bar\eta_{1,2}=0$, further simplification of the
above solutions is possible.
\subsec{Symmetry breaking and EVA mechanism}
There are five high-energy scales in this problem, which are correlated and
in principle self-consistently determined, as follows:
\item{(i)}the string unification scale $M_U\approx7.3\times g\times10^{17}\GeV$
\refs{\Kap,\Lacaze,\thresholds};
\item{(ii)}the scale of $U_A$ anomaly cancellation vevs
$\vev{\phi}\sim\sqrt{\epsilon}={\cal O}(10^{17}\GeV)$;
\item{(iii)}the $SU(5)\times U(1)_\ty\to SU(3)_C\times SU(2)_L\times U(1)_Y$
symmetry breaking scale $M_X$, undetermined theoretically but
phenomenologically
of ${\cal O}(10^{15-18}\GeV)$;
\item{(iv)}the scale where the surplus $Q,\bar Q$ get heavy $M_{eva}$
(the EVA scale \EVA), to be determined;
\item{(v)} the scales of hidden sector condensation $\Lambda_4$ and
$\Lambda_{10}$, to be determined from the hidden sector matter spectrum.
\smallskip
The gauge symmetry breaking occurs as follows
\eqna\VIx
$$\eqalignno{SU(5)\times U(1)_\ty\times U_4(1)&\supset
SU(3)_C\times SU(2)_L\times U(1)_{Y'}\times U(1)_\ty\times U_4(1)\cr
&\to SU(3)_C\times SU(2)_L\times U(1)_Y,&\VIx {}\cr}$$
where we have assumed that $U_4(1)$ remains unbroken, \ie,
$\eta_{1,2}=\bar\eta_{1,2}=0$. This appears to be a reasonable choice since
these fields do not participate in the anomalous $U_A(1)$ cancellation. Also,
if $\eta_{1,2},\bar\eta_{1,2}\not=0$, the subsequent discussion will remain
qualitatively unaffected.
We will generally assume that $V_0,V_1,\ov V_5\not=0$, $V_4=0$, and optionally
$V_{2,3},\ov V_4\not=0$. Given the charges of $\nu^c_{0,1},\bar\nu^c_5$
under the relevant $U(1)$ symmetries, it is clear that only $Y={1\over5}Y'-
{2\over5}\ty$ remains unbroken below the scale $M_X$. In fact, there are two
flat directions of the scalar potential: $\varphi_1={1\over2}(\nu^c_0+\nu^c_1)
+{1\over\sqrt{2}}\bar\nu^c_5$ and
$\varphi_2={1\over\sqrt{2}}(\nu^c_0+\nu^c_1)$.
The imaginary parts of the scalar components of these fields are eaten by the
broken $U(1)$ gauge bosons and the remaining supermultiplet components become
heavy higgs/higgsino states. There is another linearly independent field which
remains light, the so-called flaton/flatino supermultiplet \flaton. The
$d^c_{0,1},\bar d^c_5$ components of $F_{0,1},\bar F_5$ appear in the
higgs triplet mass matrix (see Subsec. 6.4). We are then left with the
$Q_{0,1},\bar Q_{4,5}$ components. The scalars get either eaten by the $X,Y$
gauge bosons or become heavy higgs bosons, whereas the fermions interact with
the $\wt X,\wt Y$ gauginos through the following mass matrix
\eqn\VIxa{{\cal M}_{1/2}=\bordermatrix{
&\bar Q_4&\bar Q_5&\wt Y\cr
Q_0&w_0^{(4)}&w_0^{(5)}&g_5V_0\cr
Q_1&w_1^{(4)}&w_1^{(5)}&g_5V_1\cr
\wt X&g_5\ov V_4&g_5\ov V_5&0\cr},}
where $w_{0,1}^{(4,5)}$ come from higher-order superpotential couplings, and
effect the EVA mechanism. We expect all these fields to become heavy, although
at different mass scales (see Sec. 7).
\subsec{The Higgs mass matrices}
The Higgs doublet masses originate from the $h_i\bar h_j\to H_i\ov H_j$
and $F_i\bar f_j\bar h_k\to V_iL_j\ov H_k$ couplings and to quartic
order are
\eqn\VIxi{{\cal M}_2=\bordermatrix{
&\ov H_1&\ov H_2&\ov H_3&\ov H_{45}\cr
H_1&0&\Phi_{12}&\bar\Phi_{31}&0\cr
H_2&\bar\Phi_{12}&0&\Phi_{23}&0\cr
H_3&\Phi_{31}&\bar\Phi_{23}&0&\bar\phi_{45}\cr
H_{45}&0&0&\phi_{45}&\Phi_3\cr
L_2&0&0&0&V_2\bar\phi_4\cr
L_3&0&0&V_3&0\cr
L_5&0&0&0&V_4\cr}.}
The Higgs triplet mass matrix which effects the doublet/triplet splitting
receives contributions from: $h_i\bar h_j\to D_i\bar D_j$,
$F_iF_jh_k\to V_iD_kd^c_j$,
$\bar F_i\bar F_j\bar h_k\to\ov V_i\bar d^c_j\bar D_k$, and
$F_i\bar F_j\to \bar d^c_j d^c_i$. The resulting matrix is
\eqn\VIxii{{\cal M}_3=\bordermatrix{
&\bar D_1&\bar D_2&\bar D_3&\bar D_{45}&d^c_0&d^c_1&d^c_2&d^c_3&d^c_4\cr
D_1&0&\Phi_{12}&\bar\Phi_{31}&0&V_1&V_0&0&0&V_4\cr
D_2&\bar\Phi_{12}&0&\Phi_{23}&0&0&0&V_2&0&0\cr
D_3&\Phi_{31}&\bar\Phi_{23}&0&\bar\phi_{45}&0&0&0&0&0\cr
D_{45}&0&0&\phi_{45}&\Phi_3&0&0&0&0&0\cr
\bar d^c_4&\ov V_4&0&0&0&w^{(4)}_0&w^{(4)}_1&0&0&\Phi_0\cr
\bar d^c_5&0&\ov V_5&0&0&w^{(5)}_0&w^{(5)}_1&0&0&\phi_3\cr}.}
\subsec{A possible numerical scenario}
Until the full set of quintic (and possibly higher-order) superpotential
couplings is calculated, one cannot work out some features of the model
such as the precise eigenvalues of the doublet and triplet mass matrices, or
the diagrams contributing to proton decay, or the details of gauge coupling
unification in this model. Nevertheless, we can postulate a possible scenario
(to be confirmed by further calculations) given the limited amount of
information available at this time.

There are three basic phenomenological considerations which help decide on the
vevs left undetermined by the flatness conditions: (i) dimension-five operators
in proton decay, (ii) the fermion Yukawa couplings and the higgs doublet
mass matrix, and (iii) the higgs triplet mass matrix. Cubic couplings of the
form $F_aF_bh\supset Q_aQ_bD$, $F_c\bar f\, \bar h\supset Q_c L \bar D$,
$h\bar h\phi\supset D\bar D\phi$ would be disastrous for proton decay \ELNpd.
We have (see Eq. \VIiii{a}): $F_0F_1h_1,F_2F_2h_2,F_4F_4h_1$;
$F_4\bar f_5 \bar h_{45},F_3\bar f_3\bar h_3$; $h_1\bar h_3\bar\Phi_{31},
h_2\bar h_3\Phi_{23}$. With our assumptions in Sec. 6.3, $Q_{0,1}$ do not
contain quark fields, whereas $Q_{2,3,4}$ do. It is then important to have
$\Phi_{23}=0$, and perhaps also $\bar\Phi_{31}=0$, assuming $Q_2(Q_4)$ contains
second (third) generation quark fields. It turns out that these two constraints
are either automatic or can be imposed (and still maintain flatness) in the
five $F$-flatness cases in Sec. 6.2. In case 1 both $H_1$ and $H_2$ become
pure massless states (see Eq. \VIxi), whereas in cases 2 and 4 (3 and 5)
$H_1$ ($H_2$) is a pure massless state. Based on the sensible assumption that
(so far neglected) higher-order contributions to ${\cal M}_2$ will likely make
massive (light) mixed states rather than (light) pure states (since the former
have more possible couplings), we conclude that cases 2 and 4 are the most
promising ones, since the others would give too many cubic fermion Yukawa
couplings.

Let us analyze the two preferred cases in turn:
\item{(a)}{\it Case 2}. We need to impose $\Phi_{23}=0$, while we get
$\bar\Phi_{31}=0$ automatically. To avoid $\ov H_{45}$ becoming heavy, we need
the $\bar\phi_{45}=0$ solution to $\phi_{45}\bar\phi_{45}=0$ (which assumes
that $\xi=\epsilon/15-{1\over2}V^2_3>0$, and is then consistent with the choice
$V_3=0$) and $V_2=0$ (also consistent with $\ov V^2_5>V^2_2$). We then get
$H_1,H_{23\ell}\propto-\vev{\Phi_{31}}H_2+\vev{\bar\Phi_{12}}H_3$, $L_{2,3,5}$;
$\ov H_2,\ov H_{45}$ light and
$H_{23h}\propto\vev{\bar\Phi_{12}}H_2+\vev{\Phi_{31}}H_3, H_{45}$; $\ov H_1,
\ov H_3$ heavy. We expect higher-order contributions to ${\cal M}_2$ to give
an intermediate scale mass to one pair of the remaining light doublets. The
cubic fermion Yukawa couplings which remain are then
$F_4\bar f_5\bar h_{45}\supset Q_4 u^c_4\ov H_{45}$ and
$F_4F_4h_1\supset Q_4 d^c_4 H_1$, leading to the identification
$\lambda_t=\lambda_b=g\sqrt{2}$. At the quartic level we have
(see Eq. \VIiii{b}) $cF_2\bar f_2\bar h_{45}\bar\phi_4$ which will give a
vanishing coupling since $\bar\phi_4=0$ in this case. Preliminary probing
into the quintic couplings reveals a structure of the form
\eqna\VIxiia
$$\eqalignno{&F_2F_2h_1\{\bar\phi^+\bar\phi^-,\bar\phi_3\bar\phi_3,
\bar\phi_4\bar\phi_4\},&\VIxiia a\cr
&\bar f_2 l^c_2h_1\{\bar\phi^+\bar\phi^-,\bar\phi_3\bar\phi_3,
\bar\phi_4\bar\phi_4\},&\VIxiia b\cr
&\bar f_5 l^c_5h_1\{\bar\phi_3\bar\phi_3,\bar\phi_4\bar\phi_4\}.&\VIxiia
c\cr}$$
All these couplings will also vanish since
$\bar\phi^+=\bar\phi^-=\bar\phi_{3,4}=0$ in this case. This scenario appears
disfavored.
\item{(b)}{\it Case 4}. Here we also need to impose $\Phi_{23}=0$, whereas
$\bar\Phi_{31}=0$ is automatic. To obtain a light $\ov H_{45}$ we need to set
$\Phi_3=V_2=V_4=0$. However, $V^2_3=2\epsilon/15>0$ in this case and the
higgs doublet mass matrix has a novel structure. We get
$H_1,H_{45},H_{23\ell}$,
$L_{2,5}$; $\ov H_2,\ov H_{45}$ light, and $H_{23h},L_3$; $\ov H_1,\ov H_3$
heavy. Note
that there will be a mixing between ``higgs" and ``lepton" doublets in this
case, since we expect one pair of the remaining light doublets to become heavy
when higher-order contributions to ${\cal M}_2$ are included. The advantage of
this case over Case 2 above is that the vevs contributing to the higher-order
couplings are generally non-zero. We obtain: $\lambda_t=\lambda_b=g\sqrt{2}$,
$\lambda_c=c\vev{\bar\phi_4}/M$, with $c={\cal O}(1)$ and
$\vev{\phi}/M\lsim1/10$ \decisive. Of the three quintic couplings in
Eq. \VIxiia{}, the first of these could give the strange-quark
Yukawa coupling, the second the tau Yukawa, and the third one the $\mu$ Yukawa.
For this assignment to be realistic we would need the second coupling
($\lambda_{\tau}$) to be sizeable despite its potential suppression by
$\vev{\phi}/M$, whereas the first coupling should remain small (and so should
the third one). These details may be possible to arrange since there are three
sources of uncertainty at this point: the actual coefficient of the terms,
the size of the singlet vevs involved, and the possibility of mixing between
$Q_2\,(d^c_2)$ and some of the other $Q$'s ($d^c$'s) in the symmetry breaking
process (higgs triplet mass matrix).
\smallskip
A study of the triplet mass matrix ${\cal M}_3$ is not very illuminating at
this time, although it is clear that we must demand $V_4=\Phi_0=\phi_3=0$
to obtain a light $d^c_4$. It is also clear that both $d^c_{0,1}$ become heavy.
In case 4, $\bar D_{1,2}; D_1, \bar d^c_5$, and a linear combination of
$D_2,D_3,\bar d^c_4$ are also heavy. Higher-order contributions to ${\cal M}_3$
are needed to continue this analysis further.
\newsec{Gauge coupling unification}
Let us now address the question which motivated our extensive search for a
5/2 model. If one starts from the low-energy values of the gauge couplings
($\sin^2\theta_w=0.2331\pm0.0013$ \EKNIII, $\alpha_3=0.113\pm0.004$ \ENR,
and $\alpha_e^{-1}=127.9\pm0.2$) and expects to get gauge coupling unification
at a scale $M_U$, then various intermediate-scale particles have to
contribute in suitable ways to the running of the gauge couplings. The
$SU(5)\times U(1)_\ty$ breaking scale $M_X$ is an unknown in the problem,
but not the only one since (as discussed in Sec. 6.3) there are other scales
inter-related to it. As an example we could imagine that $M_X$ is close to
$M_U$, in which case the expressions of Refs. \refs{\price,\SISM} apply
(for $M_X<M_U$ see Ref. \BL), as follows
\eqna\VIxiii
$$\eqalignno{L_Q&=(23.31-25.86)+\h L_{U^c}+\h L_{E^c}+2L_{\F},&\VIxiii a\cr
L_{D^c}&=(58.29-67.90)+L_L+\h L_{E^c}+2L_{\F}-\h L_{U^c},&\VIxiii b\cr}$$
where $L_R=\sum_i \ln(M_U/m_{R_i})$ and the sum runs over all the
supermultiplets in representation $R_i$. The constant ranges allow for values
of the low-energy parameters inside the 1-$\sigma$ error ellipsoid \SISM. In
the present model there are no $U^c$ or $E^c$ representations besides the
standard ones, thus $L_{U^c}=L_{E^c}=0$ in \VIxiii{}. The important point
encoded in these expressions is the fact that $L_Q>0$ and therefore
intermediate-scale extra vector-like $Q$ representations are needed. These
are not present in 4/1 models, and one pair is present in 5/2 models.
Note also that $L_{D^c}>0$ (since $L_{U^c}=0$) and we must also have extra
vector-like $D^c$ representations at intermediate mass scales, but this is
always the case in flipped $SU(5)$ models.
\smallskip
{}From Table IV and the vev choices made in Sec. 6, the contribution to the
running of the gauge couplings can be split up into five groups:
\item{1.}$F_{2,3,4},\bar f_{2,3,5},l^c_{2,3,5}$ contribute the usual three
generations of quarks and leptons.
\item{2.} $H_{1,2,3,45},\ov H_{1,2,3,45}$ contribute to
$L_L=\sum_i \ln(M_U/m^{(2)}_i)^2$, where $m^{(2)}_i$ are the eigenvalues of
${\cal M}_2{\cal M}_2^T$. (Note that $L_L$ includes the contribution from the
expected two light higgs doublets.)
\item{3.}$D_{1,2,3,45},\bar d^c_{4,5};\bar D_{1,2,3,45},d^c_{0,1}$ contribute
to $L_{D^c}=\sum_i\ln(M_U/m^{(3)}_i)^2$, where $m^{(3)}_i$ are the eigenvalues
of ${\cal M}_3{\cal M}_3^T$ with the last three columns removed from
${\cal M}_3$.
\item{4.}$Q_0,Q_1\to Q_H=(Q_0+Q_1)/\sqrt{2},Q_{eva}=(Q_0-Q_1)/\sqrt{2}$ and
$\bar Q_4,\bar Q_5\to \bar Q_H=(\ov V_4\bar Q_4+\ov V_5\bar Q_5)/\ov V,
\bar Q_{eva}=(\ov V_5\bar Q_4-\ov V_4\bar Q_5)/\ov V$, where $Q_H,\bar Q_H$
become heavy higgs/higgsino states in the $SU(5)\times U(1)_\ty$ symmetry
breaking process (see Eq. \VIxa\ with $w\ll V,\ov V$) and
$Q_{eva},\bar Q_{eva}$ get masses $M_{eva}={\cal O}(w)$. (Note that in
Eq. \VIxa\ we need $w_0^{(4)}\not=w_1^{(4)}$ or $w_0^{(5)}\not=w_1^{(5)}$ to
avoid a massless $Q_{eva},\bar Q_{eva}$ pair.) The latter contribute to
$L_Q=2\ln(M_U/M_{eva})$.
\item{5.}$\F_i,\Fb_j$ which have $\pm\h$ electric charges contribute to
$L_{\F}=\sum_i\ln(M_U/m^{(\pm1/2)}_i)^2$ where $m^{(\pm1/2)}_i$ are the
eigenvalues of the mass matrix ${\cal M}_{\F}{\cal M}_{\F}^T$ with
$\F_i{\cal M}_{\F}\Fb_j$.
\smallskip
{}From Eq. \VIxiii{a} it is clear that $L_Q\ge(23.31-25.86)$ since $L_{\F}\ge0$
(recall that $L_{U^c}=L_{E^c}=0$). From the expression for $L_Q$ just derived,
it follows that $M_{eva}\lsim10^{12}\GeV$. This upper bound is reduced to
$\approx10^{10,8,6}\GeV$ for $M_X=10^{17,16,15}\GeV$ \BL. This is the only
direct test of gauge coupling unification in this model which we can perform
at this time since it only involves the determination of the scale $M_{eva}$.
We have explored the quintic terms in the superpotential and found terms of
the form $F_{0,1}\bar F_{4,5}\vev{\phi}\vev{DD,\F\Fb}/M^2$ which would give
$w\sim M_{eva}\sim{\cal O}(10^5\GeV)$ if the $SU(4)_h$ condensation scale is
${\cal O}(10^{12}\GeV)$ and $\vev{\phi}\sim10^{17}\GeV$. These are plausible
numbers which do not violate the upper bound on $M_{eva}$ even for
$M_X=10^{15}\GeV$, and still leave room for a non-vanishing $L_{\F}$.
\newsec{Conclusions}
Gauge coupling unification is one of the few universal predictions of string
models (together with the presence of gravity and gauge interactions). An
even more pervasive fact is that the string unification scale can be precisely
calculated in any given string model. The robustness of this prediction is in
sharp constrast with the basically unlimited number of possible models with
or without supersymmetry, and with all possible gauge groups and matter
representations. (The space of models possesses certain internal structure
which for example correlates gauge groups and matter representations.)
Nevertheless, string unification in any given model can actually be tested.
Indeed, any such model gives in principle definite predictions for some of
the best measured parameters at the electroweak scale, namely $\sin^2\theta_w$
and $\alpha_3$. From this point of view, traditional unified models are also
testable this way \EKNI, although there the unification scale has to be chosen
to fit the experimental data.

The motivation for this paper was to search for a flipped $SU(5)$ model which
could possibly accommodate the measured values of the low-energy gauge
couplings
or equivalently which could unify at the string scale. There is a necessary
condition
for such a model (in level-one Kac-Moody constructions and with small string
threshold effects), that it possesses extra $Q,\bar Q$ representations,
\ie, a 5/2 model. To pursue this objective within the free fermionic
formulation, we have developed a sophisticated and comprehensive computer
program which can sweep systematically over large numbers of models and
determine their gauge group, massless spectrum, and cubic superpotential,
and decide to keep only ``interesting" models. Our search space, even though
vast, is by no means exhaustive and much of it still remains uncharted.
Nevertheless, the structure of what we did explore has proved to be very
simple.
Of the few supersymmetric 5/2 models which we found, only one of them allows
for both top- and bottom-quark Yukawa couplings. Further probe into the
superpotential and the $D$- and $F$-flatness constraints indicates that the
model is phenomenologically sound. However, much work remains to be done to
bring this model up to the level of development of its predecessor (the
revamped model). An encouraging result is that the masses of the extra
$Q,\bar Q$ states are likely to fall within the bounds imposed by string
unification.

Contrary to popular belief, string models can be tested. Moreover, if models
which satisfy {\it all} known phenomenological constraints can ever be found,
then they will become candidates for the theory of everything until they
are defeated by future tests or until string theory hands down its final
verdict, whichever comes first.
\bigskip
\bigskip
\cl{\bf Acknowledgments}\nobreak
We would like to thank I. Antoniadis and J. Ellis for useful discussions
at the earlier stages of this work.
This work has been supported in part by DOE grant DE-FG05-91-ER-40633.
The work of K.Y has been supported in part by the Texas National
Laboratory Research Commission under Grant No. RCFY9155, and in part
by the U.S. Department of Energy under Grant No. DE-FG05-84ER40141.
\listrefs
\vbox{\tenpoint\noindent {\bf Table I}: The rank and determinant \Hump\ of the
Cartan matrices $C$ of all the Lie algebras. As usual,
$A_l=SU(l+1), B_l=SO(2l+1), C_l=Sp(2l), D_l=SO(2l)$.}
\medskip
\input tables
\thicksize=1.0pt
\begintable
Algebra |${\rm det}\,C$ |rank \cr
$A_l$|$l+1$|$l$\nr
$B_l$|2|$l$\nr
$C_l$|2|$l$\nr
$D_l$|4|$l$\nr
$E_6$|3|6\nr
$E_7$|2|7\nr
$E_8$|1|8\nr
$F_4$|1|4\nr
$G_2$|1|2\endtable
\bigskip
\bigskip
\bigskip
\bigskip
\vbox{\tenpoint\noindent {\bf Table II}: The set of spin-structure basis
vectors which are common in the construction of all our models. The first
entry corresponds to the complexified $\psi^\mu$ and the next 18 to the
six left-moving triplets $(\chi^\ell,y^\ell,\omega^\ell)$. The first 12
right-moving entries (to the right of the colon) correspond to the real
fermions $\bar y^\ell,\bar\omega^\ell$, and the last 16 correspond to complex
fermions. A 1 (0) stands for periodic (antiperiodic) boundary conditions. We
also use the symbols $1_8=11111111$ and $0_8=00000000$.}
\smallskip
\hrule
$$\eqalign{
{\bf 1}&=(1\ 111\ 111\ 111\ 111\ 111\ 111\ :\ 111111\ 111111\ 11111\ 111
\ 1_8)\cr
S&=(1\ 100\ 100\ 100\ 100\ 100\ 100\ :\ 000000\ 000000\ 00000\ 000\ 0_8)
\cr
b_1&=(1\ 100\ 100\ 010\ 010\ 010\ 010\ :\ 001111\ 000000\ 11111\ 100\
0_8)\cr
b_2&=(1\ 010\ 010\ 100\ 100\ 001\ 001\ :\ 110000\ 000011\ 11111\ 010\
0_8)\cr
b_3&=(1\ 001\ 001\ 001\ 001\ 100\ 100\ :\ 000000\ 111100\ 11111\ 001\
0_8)\cr
b_4&=(1\ 100\ 100\ 010\ 001\ 001\ 010\ :\ 001001\ 000110\ 11111\ 100\
0_8)\cr
b_5&=(1\ 001\ 010\ 100\ 100\ 001\ 010\ :\ 010001\ 100010\ 11111\ 010\
0_8)\cr}$$
\hrule
\vfill\eject
\vbox{\tenpoint\noindent {\bf Table III}: Classification of $\alpha$-vectors
according to the number and type of residual $U_{a,b,c}$ gauge symmetries
which they break or do not break. $U_a:\bar w^2\bar w^3$,
$U_b:\bar y^1\bar w^6$, and $U_c:\bar y^4\bar y^5$.}
\medskip
\thicksize=1.0pt
\begintable
Class | rank | $U_a$ | $U_b$ | $U_c$ \cr
1|19|$\surd$|$\surd$|$\surd$\nr
2a|18|$\surd$|$\surd$|$\times$\nr
2b|18|$\surd$|$\times$|$\surd$\nr
2c|18|$\times$|$\surd$|$\surd$\nr
3a|17|$\surd$|$\times$|$\times$\nr
3b|17|$\times$|$\surd$|$\times$\nr
3c|17|$\times$|$\times$|$\surd$\nr
4|16|$\times$|$\times$|$\times$\endtable
\vfill\eject
\vbox{
\baselineskip=10pt
{\noindent {\bf Table IV}: The massless spectrum of the selected 5/2 model
with gauge group $SU(5)\times U(1)_\ty\times SO(10)_h\times SU(4)_h\times
U(1)^5$ and 2/3/2 Yukawa set. The transformation properties of the
observable sector fields under $SU(5)\times U(1)_\ty$ are as follows:
$F\,({\bf10},1/2)$, $\bar f\,(\bar{\bf5},-3/2)$, $l^c\,({\bf1},5/2)$,
$h\,({\bf5},-1)$. The hidden sector fields transform under
$SO(10)_h\times SU(4)_h$ as follows: $T\,({\bf10},1)$, $D\,(1,{\bf6})$,
$\F\,(1,{\bf4})$. The $\F_i,\Fb_j$ fields carry $\pm1/2$ electric charges.}
\medskip
\hrule
$$\eqalign{{\rm Observable\ Sector:}\quad
&F_0\,(-\h,0,0,-\h,0)\quad F_1\,(-\h,0,0,\h,0)\cr
&F_2\,(0,-\h,0,0,0)\quad \bar f_2\,(0,-\h,0,0,0)\quad l^c_2\,(0,-\h,0,0,0)\cr
&F_3\,(0,0,\h,0,-\h)\quad \bar f_3\,(0,0,\h,0,\h)\quad l^c_3\,(0,0,\h,0,\h)\cr
&F_4\,(-\h,0,0,0,0)\quad \bar F_4\,(\h,0,0,0,0)\cr
&\bar F_5\,(0,\h,0,0,0)\quad \bar f_5\,(0,-\h,0,0,0)\quad
                                                l^c_5\,(0,-\h,0,0,0)\cr
&h_1\,(1,0,0,0,0)\quad \bar h_1\,(-1,0,0,0,0)\cr
&h_2\,(0,1,0,0,0)\quad \bar h_2\,(0,-1,0,0,0)\cr
&h_3\,(0,0,1,0,0)\quad \bar h_3\,(0,0,-1,0,0)\cr
&h_{45}\,(-\h,-\h,0,0,0)\quad \bar h_{45}\,(\h,\h,0,0,0)\cr}$$
$$\eqalign{{\rm Singlets:}\quad
&\Phi_{12}\,(-1,1,0,0,0)\quad \bar\Phi_{12}\,(1,-1,0,0,0)\cr
&\Phi_{23}\,(0,-1,1,0,0)\quad\bar\Phi_{23}\,(0,1,-1,0,0)\cr
&\Phi_{31}\,(1,0,-1,0,0)\quad\bar\Phi_{31}\,(-1,0,1,0,0)\cr
&\phi_{45}\,(\h,\h,1,0,0)\quad \bar\phi_{45}\,(-\h,-\h,-1,0,0)\cr
&\phi^+\,(\h,-\h,0,0,1)\quad\bar\phi^+\,(-\h,\h,0,0,-1)\cr
&\phi^-\,(\h,-\h,0,0,-1)\quad\bar\phi^-\,(-\h,\h,0,0,1)\cr
&\phi_{3,4}\,(\h,-\h,0,0,0)\quad\bar\phi_{3,4}\,(-\h,\h,0,0,0)\cr
&\eta_{1,2}\,(0,0,0,1,0)\quad\bar\eta_{1,2}\,(0,0,0,-1,0)\quad
                                \Phi_{0,1,3,5}\,(0,0,0,0,0)\cr}$$
\noindent Hidden Sector:
$$\eqalign{
&T_1\,(-\h,0,\h,0,0)\quad T_2\,(-\h,-\h,0,0,-\h)\quad T_3\,(-\h,0,\h,0,0)\cr
&D_1\,(0,-\h,\h,\h,0)\quad D_2\,(0,-\h,\h,-\h,0)\quad D_3\,(-\h,0,\h,0,0)\cr
&D_4\,(-\h,-\h,0,0,\h)\quad D_5\,(0,-\h,\h,0,0)\quad D_6\,(0,\h,-\h,0,0)
\quad D_7\,(\h,0,-\h,0,0)\cr
&\F_{1}^+\,(-\q,\q,-\q,0,-\h)\quad \F_{2}^-\,(\q,\q,-\q,0,\h)
\quad \F_{3}^+\,(\q,-\q,-\q,0,\h)\cr
&\F_{4}^+\,(-\q,\tq,\q,0,0)\quad \F_{5}^+\,(-\q,\q,-\q,0,\h)
\quad\F_{6}^+\,(-\q,\q,-\q,0,-\h)\cr
&\Fb_{1}^-\,(-\q,\q,\q,\h,-\h)\quad \Fb_{2}^-\,(-\q,\q,\q,-\h,-\h)
\quad \Fb_{3}^-\,(\q,-\q,\q,0,-\h)\cr
&\Fb_{4}^-\,(-\q,\q,\q,0,-\h)\quad \Fb_{5}^+\,(-\q,-\q,\q,0,-\h)
\quad\Fb_{6}^-\,(-\tq,\q,-\q,0,0)\cr}$$
}
\vfill\eject
\bye